\PassOptionsToPackage{table,usenames,dvipsnames}{xcolor}

\documentclass[%
aps,
 reprint,
 amsmath,amssymb,
prper,
floatfix
]{revtex4-2}

\usepackage{multirow}
\usepackage{graphicx}
\usepackage{dcolumn}
\usepackage{bm}
\usepackage{tikz}
\usepackage{pgfplots}
\usepackage[export]{adjustbox}
\pgfplotsset{compat=1.17}
\usepackage{listings}
\begin{document}


\title{Multilingual Performance of a Multimodal Artificial Intelligence System on Multisubject Physics Concept Inventories}

\author{
Gerd Kortemeyer$^{1,2}$,
Marina Babayeva$^{3}$,
Giulia Polverini$^{4}$,
Ralf Widenhorn$^{5}$,
Bor Gregorcic$^{4}$
\vspace{1\baselineskip}}

\affiliation{
$^{1}$Rectorate and AI Center, ETH Zurich, 8092 Zurich, Switzerland\\
$^{2}$Michigan State University, East Lansing, MI 48823, USA\\
$^{3}$Department of Physics Education, Charles University, Prague 8, Czech Republic\\
$^{4}$Department of Physics and Astronomy, Uppsala University, 75120 Uppsala, Sweden\\
$^{5}$Department of Physics, Portland State University, Portland, OR 97207, USA
}

\date{\today}

\begin{abstract}
We investigate the multilingual and multimodal performance of a large language model-based artificial intelligence (AI) system, GPT-4o, using a diverse set of physics concept inventories spanning multiple languages and subject categories. The inventories, sourced from the PhysPort website, cover classical physics topics such as mechanics, electromagnetism, optics, and thermodynamics, as well as relativity, quantum mechanics, astronomy, mathematics, and laboratory skills. Unlike previous text-only studies, we uploaded the inventories as images to reflect what a student would see on paper, thereby assessing the system’s multimodal functionality. Our results indicate variation in performance across subjects,  with laboratory skills standing out as the weakest. We also observe differences across languages, with English and European languages showing the strongest performance. Notably, the relative difficulty of an inventory item is largely independent of the language of the test. When comparing AI results to existing literature on student performance, we find that the AI system outperforms average post-instruction undergraduate students in all subject categories except laboratory skills. Furthermore, the AI performs worse on items requiring visual interpretation of images than on those that are purely text-based. While our exploratory findings show GPT-4o's potential usefulness in physics education, they highlight the critical need for instructors to foster students' ability to critically evaluate AI outputs, adapt curricula thoughtfully in response to AI advancements, and address equity concerns associated with AI integration.
\end{abstract}

\maketitle

\section{Introduction}

\subsection{Generative artificial intelligence in physics education}
The public availability of Large Language Models (LLMs), like those built on the architecture introduced by Vaswani et al.~\cite{vaswani2017attention}, has unlocked new possibilities across various domains, including education~\cite{kung2022,lawexam}. Since the release of ChatGPT in the fall of 2022~\cite{chatgpt}, LLMs have surged in popularity, with scholars showcasing their remarkable capabilities. Beyond the wave of enthusiasm generated by human-like responses that have been shown to pass the Turing Test~\cite{turing1950} with a majority of human test subjects~\cite{jones2024people}, the OpenAI's GPT series demonstrated proficiency in academic fields such as physics in a number of benchmarks~\cite{achiam2023gpt}. Both the initial version and later iterations, particularly GPT-4~\cite{gpt4}, have achieved impressive results in physics, such as passing standardized exams, excelling in introductory courses and even coming close to passing entire degrees~\cite{kortemeyer23ai,polverini2024ejp,kortemeyer24cheating,yeadon2024impact,pimbblet_2024}.

The technology is increasingly being embraced in physics education~\cite{sperling2024artificial}, offering promising applications for both teaching and learning physics. LLMs have proven valuable for teachers, helping in the creation of tailored materials and tasks, student assessments~\cite{kuchemann2024chatgpt}, and personalized feedback~\cite{bitzenbauer2023,kuchemann23,kortemeyer2023using, wan24, chen2025,fussell2025comparing}, as well as for their training~\cite{gregorcic24}. For students, these models represent ever-available, patient, and knowledgeable resources~\cite{crawford2023,vasconcelos2023}. However, these opportunities come with significant risks, particularly the potential for users to overly trust AI when assessing its scientific accuracy~\cite{dahlkemper23,ding2023students}.

Early publicly available generative AI systems have been limited to processing and generating text-based content only. Thus, earlier studies of AI's performance on physics tasks, such as concept inventories, were restricted to text-based materials or textual descriptions of visual elements~\cite{kortemeyer23ai,west2023ai,wheeler2023chatgpt,cho2024investigation,aldazharova2024assessing}. More recently, multimodal systems, which can also input and output auditory and visual data, have broadened the scope of such studies (e.g.,~\cite{polverini2024evaluating}). 

In this study, we confront GPT-4o~\cite{gpt4o}, a popular AI model from OpenAI, with screenshots of tasks from physics concept inventories, which reflect the test items as seen by learners, including the accompanying diagrams, sketches, graphs, and illustrations. Earlier studies using GPT-4 and GPT-4o with the FCI~\cite{hestenes1992}, TUG-K~\cite{beichner1994testing} and BEMA~\cite{ding2006evaluating} concept inventories demonstrated promising behavior, but also significant challenges, mainly related to the AI system's limited ability to interpret visual information~\cite{aldazharova2024assessing,polverini24,polverini2024performance}. We considered the English version of each selected concept inventory and all its available translations.

\subsection{Concept inventories in physics}
Examinations in physics generally involve symbolic and numerical calculations, whereas concept inventories are typically different, focusing primarily on conceptual understanding~\cite{smith2010problem,sands2018using}. Scores on these different types of assessments do not necessarily correlate, as for example, the FCI would under-predict success in a calculus-based physics course~\cite{henderson2002common}. On the surface, this would work in AI's favor, as AI systems used to be notoriously ``bad at math,'' which hampered their performance on traditional physics exam questions~\cite{kortemeyer23ai}; this is remedied in newer models which generate Python code for calculations or are explicitly ``reasoning,'' such as GPT-o1 and GPT-o3-mini~\cite{gpto1,geisler2025}. However, early investigations of AI's performance on conceptual tasks suggest that these, too, can present several challenges~\cite{gregorcic23,polverini2024ejp}.

Concept inventories have played and continue to play an important role in physics education research~\cite{madsen2017best}, and some of the most influential studies have been based on their outcomes, most notably with respect to active engagement~\cite{hake1998interactive}. Unlike traditional assessments that focus on individual learners, concept inventories are primarily designed to evaluate instructional methods, oftentimes with a focus on learning gains rather than the absolute scores. As we embark on assessing AI on the base of absolute scores, we deviate from this practice.

Arguably, while many concept inventory items assess students' understanding of core ideas and concepts~\cite{sands2018using}, they do not necessarily capture evidence of scientific practices or crosscutting concepts~\cite{laverty2018analysis} (even though this distinction is debatable~\cite{stoen2020force}). It is thus important to emphasize that our study assesses AI's performance on physics conceptual tasks, but does not evaluate if it has the qualities of a physicist.

\subsection{The language problem}
Language plays a crucial role in learning physics. Expert physicists structure their knowledge using layered metaphorical systems and specific grammatical frameworks, which shape how they communicate complex ideas~\cite{brookes2007using,euler2019embodiment}. These unconscious linguistic patterns can influence how students understand physics~\cite{brookes2006role,brookes2007using}. Although English has become the \textit{lingua franca} of most scientific communications, undergraduate students tend to be instructed in their native language, which is also the language in which they will tend to tackle physics problems.

LLMs hold the promise of facilitating language-related learning tasks~\cite{wulff2024physics}, yet they currently do not function equally well in all languages; for example, OpenAI's research on the GPT-4's performance  shows inconsistent results across different languages~\cite{gpt4research}. Due to the disparities in the prevalence, quantity, and quality of information available across languages, there exists a disparity in the resources available for LLM training, which can have an impact on model performance. Nicholas and Bhatia highlight that although LLMs are designed to mitigate the issue of underrepresentation of certain languages in learning data, as of 2023, early LLMs were still predominantly trained on English materials~\cite{nicholas2023lost}. While this may be changing as governments and companies attempt to strengthen AI capabilities in their countries (e.g., models like DeepSeek~\cite{Deepseek} or Qwen~\cite{Qwen} developed by companies in China, or the Swiss AI Initiative~\cite{swissai}), it is likely that LLMs remain biased toward the needs of major economies with the financial resources needed to train AI systems.  Similar issues are also discussed in the analysis conducted by ``Cohere for AI'' company in their report titled ``The AI Language Gap''~\cite{cohere}. Feng et al. also acknowledge the disparity in LLM performance across different languages when it comes to abstaining from hallucinations, resulting in a gap of approximately 20\% between high-resource and low-resource languages~\cite{feng2024teaching}; ``hallucinations'' refer to instances where an  LLM generates plausible-sounding but inaccurate or entirely fabricated information, which are indicative of shortcomings in calibration and reasoning. Particularly relevant for our study is a recent finding that a Chinese-trained model performed better on the FCI when prompted in Chinese rather than English~\cite{cho2024investigation}.

Understanding and using physics-specific language is essential for physics literacy and learning. To address both the linguistic disparities of LLMs and the specialized nature of physics language, it is important to examine how LLMs perform across diverse languages in the context of physics education. While current research highlights the general challenges of multilingual use of LLMs, the intersection of language and subject-specific terminology, such as physics, is understudied and lacks a clear understanding of the current situation.

\subsection{Relevance to Physics Education Research}\label{sec:relevant}
While AI’s potential as a learning tool, assessment assistant, and research aid is widely acknowledged~\cite{kotsis24, yeadon2024impact, tschisgale23,tschisgale23err,odden24}, the performance of LLMs on established and validated physics conceptual assessments, particularly in multilingual contexts, and on multiple subject domains, remains underexplored. 

As AI assumes an increasingly large role in the educational process, it is also important for physics education stakeholders to develop a sense of its capabilities in physics-related tasks. Assessing AI's ability to solve the kind of problems we use for assessing student understanding of physics concepts is a necessary step if we want to make appropriate and responsible use of these systems in physics education. The physics education research community, with its wealth of research-based assessment instruments, so-called concept inventories, is well positioned to engage in such evaluation. Concept inventories are developed to be robust tools for assessing university students' conceptual understanding. Comparing AI's performance to that of students can thus provide a student-centered reference point that is more meaningful for PER researchers than AI-facing benchmark assessments (which are oftentimes designed to assess physical reasoning tasks and inform machine-learning engineering~\cite{bakhtin2019phyre,xue2021phy,achiam2023gpt,melnik2023benchmarks,melnik2023benchmarks,physbench2025}), and they are more standardized and universal than exams from individual courses (e.g.,~\cite{pimbblet_2024}). 

Comparing AI's average performance to university students' post-instruction performance can thus provide a rough student-centric measure of the level of performance of AI systems. There is a need for caution here, however. There are important differences in information processing in students and AI models. This means that, for example, when AI reaches a numerical performance similar to that of an average student, it does not necessarily mean that its strengths and difficulties are similar to those of an average student. In fact, previous studies have shown important differences in the profile of students' and AI's difficulties --- i.e. AI's displayed difficulties have been found to be uncharacteristic of typical student difficulties ~\cite{polverini2024performance, polverini24}.

One of the motivations for our work is exploring what types of tasks might be difficult for GPT-4o to solve. 
For instructors, knowing this would allow them to communicate to their students the strengths and weaknesses of the AI system, so that students may use it in responsible and productive ways when working unsupervised (e.g., while doing homework or project assignments) or in contexts beyond formal education.
Exploring the different facets of an AI's performance can also inform developers about AI’s potential as an educational tool for automated grading ~\cite{kortemeyer2024assessing, mok25, chen2025}, generating feedback~\cite{guo2024, krupp2024}, and personalizing instruction~\cite{agui2022}. However, for AI to be genuinely useful in these applications, it must demonstrate consistency and reliability across diverse educational contexts. For example, a recent study suggests that an AI system's ability to grade student answers on a topic is correlated to its problem-solving ability~\cite{mok25} on that topic. Analyzing GPT-4o’s performance on structured, standardized conceptual assessments, can therefore also provide insights into whether AI can reliably assist instructors in evaluating student understanding and be integrated into automated feedback and grading systems.

Another major motivation driving this research is the impacts of AI’s increasing accessibility on student learning and assessment validity in physics education~\cite{kortemeyer24cheating}. With generative AI now readily available, students are already using it for study purposes, sometimes in ways that challenge traditional expectations of academic integrity~\cite{lee2024}.  More broadly, this issue ties into an ongoing discussion in PER regarding how assessments need to evolve in an AI-assisted learning environment and how physics curricula should adapt to the fact that many tasks can now be outsourced to AI.

Another key aspect of this study is its multilingual approach. While much of PER is conducted in English~\cite{docktor2014synthesis,meltzer2015brief}, physics is taught in a wide range of languages, and the educational impact of AI should be considered beyond English-speaking classrooms. LMMs like GPT-4o have been trained primarily on English-language data~\cite{nicholas2023lost}, which raises concerns about whether their accuracy is consistent across different linguistic versions of physics assessments. If AI models perform significantly better in one language than in another, this could reinforce existing inequities in educational technology~\cite{bulathwela24}. Understanding these potential disparities is necessary to ensure equitable access to AI-assisted learning tools and to assess whether AI can serve as a reliable resource for students who learn physics in languages other than English. 

Lastly, our study contributes to a growing body of research on the multimodal performance of LLMs, particularly on tasks involving physics visual representations. This ties into the question of what types of tasks are especially difficult for AI to solve. While it is known that GPT-4o struggles with tasks involving kinematics graphs \cite{polverini2024evaluating}, it remains unclear if it experiences similar limitations across other types of physics problems and subject categories, and whether these weaknesses persist when the visual data are embedded in diverse real-world assessment formats. 

This study is exploratory in nature; rather than aiming to provide definitive answers to broad questions about AI’s role in physics education, it seeks to identify emerging trends in AI performance on physics concept inventories. By analyzing GPT-4o’s accuracy on multiple-choice questions across different languages, with and without accompanying images, we aim to highlight patterns that warrant further investigation rather than making strong claims about AI’s conceptual reasoning abilities. Any broader generalizations about AI’s impact on assessment, instruction, or research would require a more detailed analysis of its reasoning processes and interactions with learners, which is beyond the scope of this paper. Nevertheless, by mapping out these initial trends, this study contributes to a growing conversation in PER about how AI is reshaping physics education, helping to inform future research directions and the responsible development and integration of AI into learning environments and educational processes.

\subsection{Research questions}
Despite a growing number of exploratory studies on LLMs in physics education~\cite{liang2023exploring,kuchemann2024chatgpt,bitzenbauer2023,kuchemann23, wan24, chen2025,fussell2025comparing,aldazharova2024assessing,dahlkemper23,ding2023students,yeadon2024impact,latif2024physicsassistant,lieb2024student,kahaleh2025evaluating,avila2024using,zhu2024personalised,kahaleh2025evaluating}, including some of our own~\cite{kortemeyer23ai, polverini2024performance, polverini2024evaluating}, there remains a lack of a broad, systematic analysis of how multimodal AI models perform on validated, research-based physics assessments, particularly when these assessments include diverse subject domains, visual components, and multiple languages. Prior studies have often focused on individual inventories or English-only text-based inputs. This study extends that work by (i) synthesizing performance trends across inventories and physics subject categories, (ii) exploring performance in different languages, (iii) comparing AI performance to student benchmarks, and (iv) reevaluating AI's visual interpretation challenges in a multilingual and multisubject framework. The following research questions guide our exploratory analysis:
\begin{itemize}
\item RQ1. \textit{How does GPT-4o perform across different physics concept inventories in English?} 
\item RQ2. \textit{How does language influence the performance of the AI system?}
\item RQ3. \textit{How does its performance compare to student performance at the undergraduate level?} 
\item RQ4. \textit{How does the presence of images influence the performance of the AI system?}
\end{itemize}
These questions are, in turn, addressed in Section IV.

\section{Data set}
\begin{figure*}
\begin{center}
\includegraphics[width=0.495\textwidth]{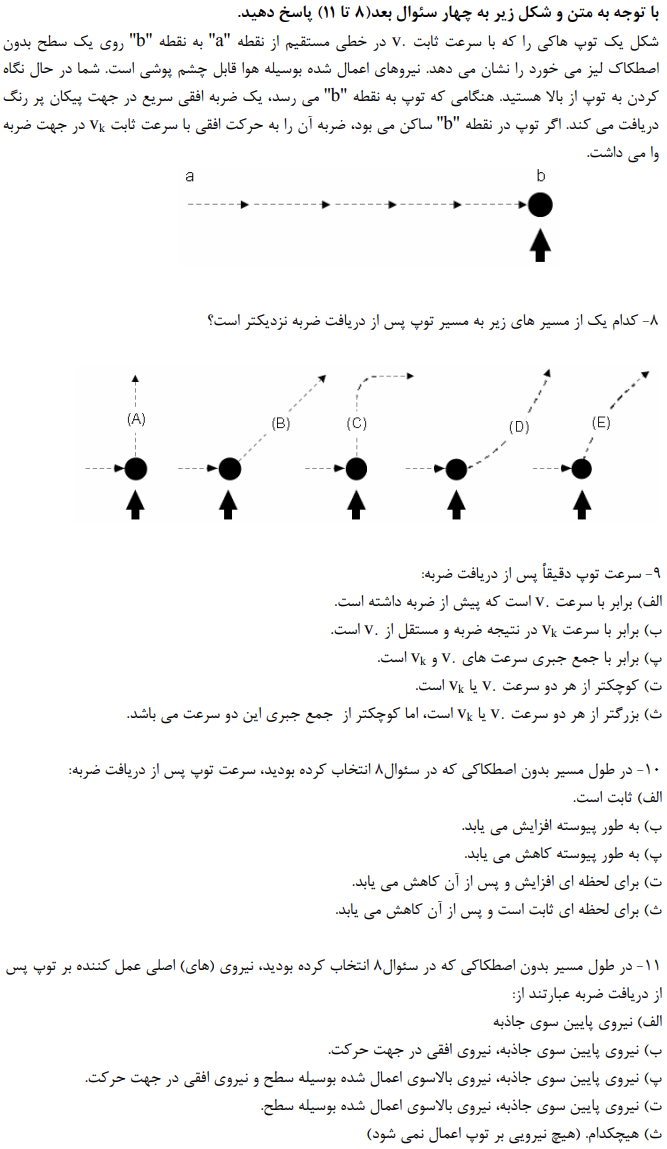}
\includegraphics[width=0.495\textwidth]{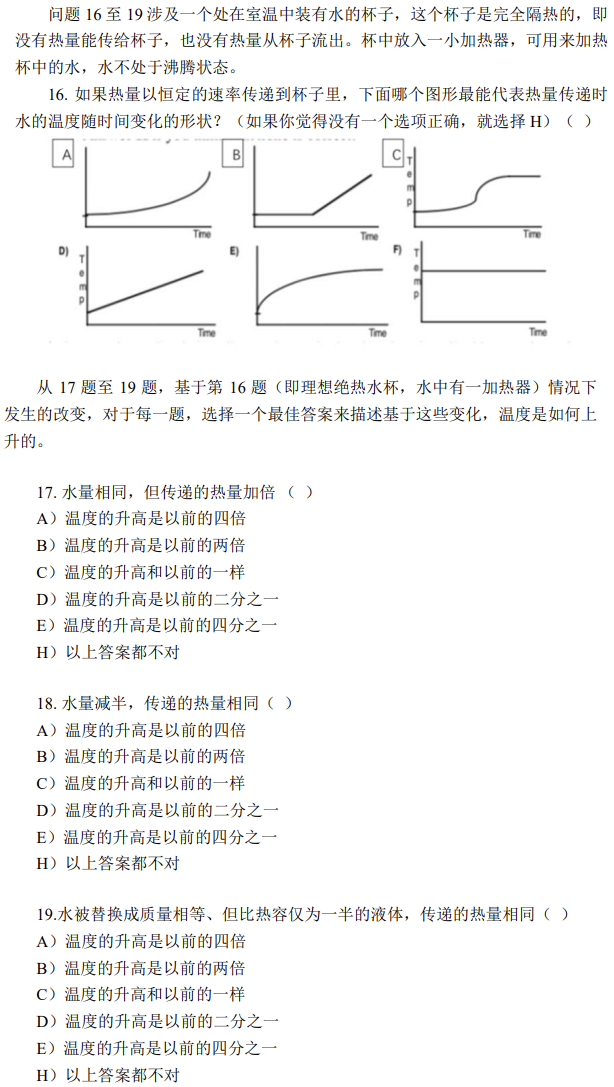}
\end{center}
\caption{Examples of uploaded problem images: FCI, items 8-11, in Persian (left panel) and HTCE, items 16-19, in Chinese (right panel).}
\label{fig:examples}
\end{figure*}

As our dataset, we used concept inventories published in PhysPort~\cite{physport}, containing a comprehensive collection of research-based inventories in multiple translations~\cite{mckagan2020physport}. We included inventories that had at least a ``bronze star" classification assigned on PhysPort. This indicates that they had been studied with respect to at least three of the seven research validation categories that the platform employs (research into student thinking; studied with student interviews, expert review, and appropriate statistical analysis; research conducted at multiple institutions, by multiple research groups, and with peer-reviewed publication). We excluded inventories for which we could not obtain an answer key. Figure~\ref{fig:examples} shows inventory items from the Force Concept Inventory in Persian and the Heat and Temperature Conceptual Evaluation in Chinese. These inventories cover a broad spectrum of physics subject categories, listed in Table~\ref{tab:sub_cat_used}.

Thirty-five languages are currently represented in PhysPort. While the authors are familiar with some of the languages, the quality of the majority of the non-English translations could not be evaluated.

Tables~\ref{tab:inventories1} through~\ref{tab:inventories4} show the concept inventories included in our investigation, along with their available translations and literature references. For some inventories, more than one version was available. The column ``\%Post'' lists some post-instruction inventory scores reported in the literature for undergraduate-level courses; these are collected best-effort and not necessarily representative (typically and much more systematically, inventory gains are reported~\cite{von2016secondary}). We omitted results with very small sample sizes, as well as reported scores at the graduate-student or post-doctoral level. For subsequent comparisons, the scores found for each inventory were averaged. 

The rightmost column ``Cat.'' shows the subject-category identifier under which we classified the test. Explanations for the abbreviations used in this column are provided in Table~\ref{tab:sub_cat_used}. These labels are derived from the PhysPort classification, however, we divided electricity and magnetism into EM-F (primarily dealing with fields and potentials), and EM-C (primarily focusing on DC and AC circuits). The multi-subject Next Gen Physical Science Diagnostic (NGPSD) was given no classification. 

\begin{table}
\caption{Subject categories of the concept inventories we investigated, as well as their abbreviations used throughout.}

\begin{ruledtabular}
\begin{tabular}{ll}
Abbreviation&Description\\
\hline
AST	& Astronomy \\
EM-F & Electricity/Magnetism - Fields	\\
EM-C & Electricity/Magnetism - Circuits	\\
MATH	& Mathematics \\
LAB	& Laboratory Skills\\
MECH	& Mechanics\\
OPT	& Optics\\
QP	& Quantum Physics\\
REAS	& Reasoning\\
RELA	& Relativity\\
THERM	& Thermodynamics
\label{tab:sub_cat_used}
\end{tabular}
\end{ruledtabular}

\label{tab:cats}
\end{table}

\begin{table*}
\caption{Concept inventories under consideration. Descriptions are taken from PhysPort~\cite{mckagan2020physport,physport}.}
\begin{ruledtabular}
\setlength{\extrarowheight}{0.5mm}
\begin{tabular}{l>{\raggedright\arraybackslash}p{3.3cm}p{4.8cm}lp{0.9cm}p{1cm}p{4cm}l}
Title&Full Title&Description&Refs.&Vers.&\%Post&Languages&Cat.\\
\hline
ADT	&Astronomy Diagnostic Test & Astronomy content knowledge (apparent motion of the sun, scale of the solar system, phases of the moon, linear distance scales, seasons, global warming, nature of light, gravity, stars, cosmology)&\cite{hufnagel2002development}& 2.0 & 
41~\cite{brogt2007analysis}\newline
54~\cite{brogt2007analysis}
&
English, Spanish, Swedish&AST\\
BEMA&Brief Electricity and Magnetism Assessment & Electricity/Magnetism content knowledge (circuits, electrostatics, magnetic fields and forces)&\cite{ding2006evaluating}&1& 
42~\cite{ding2006evaluating}\newline
43~\cite{koca2024evaluating}\newline
61~\cite{pollock2008comparing}
&
Chinese, English, Japanese, Portuguese, Spanish, Swedish &EM-F\\
CCI&Calculus Concept Inventory&Mathematics content knowledge (functions, derivatives, limits, ratios, the continuum)&\cite{epstein2006calculus}&5&
50~\cite{maciejewski2016flipping}\newline
52~\cite{maciejewski2016flipping}
&Czech, English&MATH\\
CDPA&Concise Data Processing Assessment&Lab skills (uncertainty in measurement, relationship between functions graphs and numbers)&\cite{day2011development}&2& 
39~\cite{day2011development}
&English, Spanish&LAB\\
CSEM&Conceptual Survey of Electricity and Magnetism&Electricity / Magnetism content knowledge (electrostatics, magnetic fields and forces, Faraday's law)&\cite{maloney2001surveying}&H&
61~\cite{tapping2019visualizing}\newline
66~\cite{pollock2008comparing}
&English,  Indonesian, Malay, Spanish, Swedish&EM-F\\
CTSR&Lawson Classroom Test of Scientific Reasoning&	Scientific reasoning (proportional thinking, probabilistic thinking, correlational thinking, hypothetico-deductive reasoning)&\cite{lawson1978development}&2&
54~\cite{moore2012scientific}\newline
75~\cite{moore2012scientific}
&English, Spanish, Swedish&REAS\\
DIRECT&Determining and Interpreting Resistive Electric Circuit Concepts Test&Electricity / Magnetism content knowledge (DC circuits)&\cite{engelhardt2004students}&1.2&
44~\cite{engelhardt2004students}\newline
63~\cite{sangam2012conceptual}
&Chinese, English, Finnish, German, Greek, Spanish, Swedish &EM-C\\
DS&Density Survey&Mechanics content knowledge (density)&\cite{yeend2001student}&1&
57~\cite{zenger2022exploring}
& English, German&MECH\\
ECA&Energy Concept Assessment&Mechanics content knowledge (energy principle, forms of energy, work and heat, absorption/emission spectrum, specifying appropriate systems)&\cite{ding2013students}&2&
47~\cite{ding2013students}\newline
50~\cite{ding2013students}
&Croatian, English &MECH\\
ECCE&Electric Circuits Conceptual Evaluation&Electricity / Magnetism content knowledge (DC and AC circuits)&\cite{sokoloff1996teaching}&1&
37~\cite{kortemeyer2019using}\newline
42~\cite{sokoloff1996teaching}\newline
64~\cite{sokoloff1996teaching}
&English&EM-C\\
EMCA&Electricity and Magnetism Conceptual Assessment&Electricity / Magnetism content knowledge (electrostatics, electric fields and force, circuits, magnetism, induction)&\cite{mccolgan2017assessing}&1&
49~\cite{mccolgan2017assessing}\newline
58~\cite{mccolgan2017assessing}
&English, Indonesian&EM-F\\
EMCS&Energy and Momentum Conceptual Survey&Mechanics content knowledge (energy, momentum)&\cite{singh03multiple}&1&
49~\cite{singh03multiple}\newline
52~\cite{singh03multiple}\newline
54~\cite{sahin2010impact}\newline
69~\cite{sahin2010impact}
&English, Finnish, Indonesian, Swedish&MECH\\
FCI&Force Concept Inventory&Mechanics content knowledge (forces, kinematics)&\cite{hestenes1992}&v95&
38~\cite{mason2020learning}\newline
56~\cite{kortemeyer09}\newline
66~\cite{han15}
&
Arabic,	Bengali, Catalan, Chinese, Croatian, Czech, Dutch, English, Filipino, Finnish, French, German, Greek, 	Hebrew,	Hindi,	Hungarian, Icelandic,	Italian,	Japanese,  Malay,	Norwegian,	Persian,	Polish,	Portuguese,	Punjabi, Russian, Slovak, Spanish, Swedish, Tamil,	Thai, Turkish&MECH\\
FMCE&Force and Motion Conceptual Evaluation&Mechanics content knowledge (kinematics, forces, energy, graphing)&\cite{thornton1998assessing}&v99&
55~\cite{cummings1999evaluating}
&English, Indonesian, Japanese, Spanish&MECH\\

\end{tabular}
\end{ruledtabular}
\label{tab:inventories1}
\end{table*}

\begin{table*}
\caption{Concept inventories under consideration (cont.). Descriptions are taken from PhysPort~\cite{mckagan2020physport,physport}.}
\begin{ruledtabular}
\setlength{\extrarowheight}{0.5mm}
\begin{tabular}{l>{\raggedright\arraybackslash}p{3.4cm}p{5.6cm}lp{1cm}p{1.5cm}p{2.5cm}l}
Title&Full Title&Description&Refs.&Vers.&\%Post&Languages&Cat.\\
\hline
FORT&Montana State University Formal Reasoning Test&Scientific reasoning (hypothesis testing, correlational reasoning, probability, control of variables, proportional reasoning)&\cite{kalinowski2019development}&1&
55~\cite{kalinowski2019development}
&English&REAS\\
FTGOT&Four-tier Geometrical Optics Test&Waves / Optics content knowledge (plane mirrors, spherical mirrors, lenses); used as two-tier test&\cite{kaltakci2017development}&1&
18~\cite{kaltakci2017development}
&English, Turkish&OPT\\
FVA&Force, Velocity, and Acceleration Test&Mechanics content knowledge (forces, velocity, acceleration)&\cite{rosenblatt2011systematic}&3.2.3a&
38~\cite{rosenblatt2011systematic}\newline
70~\cite{rosenblatt2011systematic}
&English&MECH\\
GECI&Greenhouse Effect Concept Inventory&Astronomy content knowledge (types of greenhouse gases, types of electromagnetic energy, energy equilibrium balance, greenhouse effect mechanisms, global warming vs. greenhouse effect)&\cite{keller2006part}&vC&
55~\cite{keller2006part}
&English, Japanese&AST\\
HTCE&Heat and Temperature Conceptual Evaluation&Thermal / Statistical content knowledge (temperature, phase change, heat transfer, thermal properties of materials)&\cite{tanahoung2006surveying}&1&
78~\cite{tanahoung2006surveying}
&Chinese, English&THERM\\
IBCDC&Inventory of Basic Conceptions -- DC Circuits&Electricity / Magnetism content knowledge (DC circuits)&\cite{halloun2007evaluation}&F06&
English:~40\newline
French:~35 \cite{halloun2007evaluation}
&English, French&EM-C\\
IBCM&Inventory of Basic Conceptions -- Mechanics&Mechanics content knowledge (forces, kinematics)&\cite{halloun2007evaluation}&F06&
English:~26 \newline
French:~31 \cite{halloun2007evaluation}
&English, French&MECH\\
LPCA&Light Phenomena Conceptual Assessment&Waves / Optics content knowledge (reflection, refraction, Snells law, wavelength and frequency, light scattering, electromagnetic spectrum, the human eye)&\cite{ndihokubwayo2020light}&1&
41~\cite{ndihokubwayo2020light}
&English&OPT\\
LPCI&Lunar Phases Concept Inventory&Astronomy content knowledge (phases of the moon)&\cite{lindell2002developing}&3&
42~\cite{lindell2002developing}\newline
55~\cite{lindell2002developing}
&English, Spanish&AST\\
LSCI&Light and Spectroscopy Concept Inventory&Astronomy content knowledge (light, waves, spectroscopy)&\cite{bardar2007development}&1&
47~\cite{wallace2018item}\newline
51~\cite{wallace2018item}\newline
52~\cite{wallace2018item}
&English&AST\\
MBT&Mechanics Baseline Test&Mechanics content knowledge (kinematics, forces, momentum, energy)&\cite{hestenes1992mechanics}&97&
35~\cite{millan2021thirty}\newline
48~\cite{antwi2011impact}\newline
66~\cite{kadar2019knowledge}\newline
73~\cite{kadar2019knowledge}
&English, Finnish, French, German, Greek,      Italian, Japanese, Malay,    Persian, Portuguese, Spanish, Turkish &MECH\\
MCS&Magnetism Conceptual Survey&Electricity / Magnetism content knowledge (magnetic fields and forces, Faraday's law)&\cite{li2016developing}&1&
41~\cite{li2016developing}\newline
44~\cite{li2016developing}
&English&EM-F\\
MUQ&Measurement Uncertainty Quiz&Lab skills (calculating error from measurements, accuracy and precision, sources of error)&\cite{deardorff2001introductory}&1&
---
&English&LAB\\
\multirow{2}{*}{MWCS} &
\multirow{2}{3.4cm}{Mechanical Wave Conceptual Survey} &
\multirow{2}{6cm}{Waves / Optics content knowledge (mechanical waves, wave propagation, wave superposition, wave reflection, standing waves)} &
\multirow{2}{*}{\cite{tongchai2009developing}} &
1 & 
\multirow{2}{*}{47~\cite{santoso2022principal}}
&English, Spanish, Thai  &
\multirow{2}{*}{OPT} \\
&&&&2& & English, Spanish& \\ [22pt]
NGCI&Newtonian Gravity Concept Inventory&Astronomy content knowledge (directionality of gravity, force law, thresholds related to gravity, independence of forces)&\cite{williamson2013development}&3&
50~\cite{williamson2013development}\newline
55~\cite{williamson2013development}
&Arabic, English&AST\\
NGPSD&Next Gen Physical Science Diagnostic&Mechanics content knowledge (magnetism, static electricity, energy, forces, waves and sound, light)&\cite{engelhardt2018developing}&2&
---
&English&---\\

\end{tabular}
\end{ruledtabular}
\label{tab:inventories2}
\end{table*}

\begin{table*}
\caption{Concept inventories under consideration (cont.). Descriptions are taken from PhysPort~\cite{mckagan2020physport,physport}.}
\begin{ruledtabular}
\setlength{\extrarowheight}{0.5mm}
\begin{tabular}{l>{\raggedright\arraybackslash}p{3.6cm}p{6cm}lp{1cm}p{1cm}p{2cm}l}
Title&Full Title&Description&Refs.&Vers.&\%Post&Languages&Cat.\\
\hline

PIQL&Physics Inventory of Quantitative Literacy&Mathematics scientific reasoning (proportional reasoning, reasoning with signed quantities, co-variational reasoning)&\cite{white2021physics}&2.2&
55~\cite{white2021physics}
&English&REAS\\

\multirow{2}{*}{QMCA} &
\multirow{2}{3.4cm}{Quantum Mechanics Concept Assessment} &
\multirow{2}{6cm}{Modern / Quantum Content knowledge (wave functions, measurement, time dependence, probability, infinite square well, 1D tunneling, energy levels, spins)} &
\multirow{2}{*}{\cite{sadaghiani2015quantum}} & 
5.5.7 & 
\multirow{2}{*}{54~\cite{sadaghiani2015quantum}}
&English, Portuguese  &
\multirow{2}{*}{QP} \\
&&&&6.6.2& &English & \\[22pt]

QMCS&Quantum Mechanics Conceptual Survey&Modern / Quantum content knowledge (wave functions, probability, wave-particle duality, uncertainty principle, infinite square well, one-dimensional tunneling, energy levels)&
\cite{mckagan2010design}&2.0&
51~\cite{mckagan2010design}\newline
69~\cite{mckagan2010design}
&English, Finnish, Japanese & QP\\
QMFPS&Quantum Mechanics Formalism and Postulates Survey&Modern / Quantum content knowledge (quantum mechanics formalism, quantum mechanics postulates)&\cite{marshman2019validation}&29&
32~\cite{marshman2015improving}\newline
37~\cite{marshman2015improving}
&English, Spanish&QP\\
QMS&Quantum Mechanics Survey&Modern / Quantum content knowledge (wave functions, measurement, expectation values, Hamiltonian, time dependence, probability, infinite square well, finite square well, harmonic oscillator, 1D tunneling)&
\cite{zhu2012surveying}&18&
38~\cite{zhu2012surveying}
&English&QP\\
QMVI&Quantum Mechanics Visualization Instrument&Modern / Quantum content knowledge (wave functions, probability, infinite square well, 1D tunneling, time dependence, momentum space, 2D potentials, visualization of the relationship between potentials and wave functions)&
\cite{cataloglu2002testing}&0.4&
28~\cite{cataloglu2002testing}\newline
29~\cite{cataloglu2002testing}\newline
45~\cite{cataloglu2002testing}\newline
58~\cite{cataloglu2002testing}
&English&QP\\
QPCS&Quantum Physics Conceptual Survey&Modern / Quantum content knowledge (de Broglie wavelength, double slit interference, uncertainty principle, photoelectric effect, wave particle duality)&\cite{wuttiprom2009development}&1&
59~\cite{wuttiprom2009development}\newline
75~\cite{wuttiprom2009development}
&English, Thai&QP\\
RAPT&Rate and Potential Test&Electricity / Magnetism Content knowledge (electric potential, rate of change)&\cite{allain2001investigating}&1A&
61~\cite{allain2001investigating}\newline
73~\cite{allain2001investigating}
&English&EM-F\\
RCI&Relativity Concept Inventory&Modern / Quantum content knowledge&\cite{aslanides2013relativity}&1&
71~\cite{aslanides2013relativity}
&English&RELA\\
RFCI&Representational Variant of the Force Concept Inventory&Mechanics content knowledge (kinematics, forces, graphing, multiple representations)&\cite{nieminen2010force}&2007&
60~\cite{nieminen2010force}\newline
61~\cite{nieminen2010force}
&English, Finnish&MECH\\
RKI&Rotational Kinematics Inventory&Mechanics content knowledge (Part 1: rotational kinematics of a particle, Part 2: rotational kinematics of a particle in rectilinear motion, Part 3: rotational kinematics of a rigid body about a fixed axis)&
\cite{mashood2012inventory}&1&
38~\cite{suarez2023tutorials}\newline
55~\cite{suarez2023tutorials}
&English, French&MECH\\
RRMCS&Rotational and Rolling Motion Conceptual Survey&Mechanics content knowledge (rotational kinetic energy, torque, rotational kinematics, moment of inertia)&\cite{rimoldini2005student}&1&
75~\cite{rimoldini2005student}
&English&MECH\\
SGCE&Symmetry and Gauss's Law Conceptual Evaluation&Electricity / Magnetism content knowledge (symmetry, electric field, electric flux)&\cite{singh2006student}&1&
49~\cite{singh2006student}
&English&EM-F\\
SPCI&Star Properties Concept Inventory&Astronomy content knowledge (stellar properties, nuclear fusion, star formation)&\cite{bailey2012development}&4&
30~\cite{bailey2012development}\newline
51~\cite{bailey2012development}
&English,  Japanese, Spanish&AST\\

\end{tabular}
\end{ruledtabular}
\label{tab:inventories3}
\end{table*}

\begin{table*}
\caption{Concept inventories under consideration (cont.). Descriptions are taken from PhysPort~\cite{mckagan2020physport,physport}.}
\begin{ruledtabular}
\setlength{\extrarowheight}{0.5mm}
\begin{tabular}{l>{\raggedright\arraybackslash}p{3.4cm}p{5cm}lp{1cm}p{1.1cm}p{3cm}l}
Title&Full Title&Description&Refs.&Vers.&\%Post&Languages&Cat.\\
\hline
\multirow{2}{*}{STPFaSL} &
\multirow{2}{3.4cm}{Survey of Thermodynamic Processes and First and Second Laws} &
\multirow{2}{5cm}{Thermal / Statistical content knowledge (first law of thermodynamics, second law of thermodynamics, PV diagrams, reversible processes, irreversible processes)} &
\multirow{2}{*}{\cite{brown2021development}} &
short& 
\multirow{2}{*}{37~\cite{brown2021development}}
&Chinese, English, Indonesian, Portuguese   &
\multirow{2}{*}{THERM} \\
&&&&long& &English & \\[42pt]

TCE&Thermal Concept Evaluation&Thermal / Statistical content knowledge (temperature, heat transfer, phase change, thermal properties of materials)&\cite{yeo2001introductory}&1&
78~\cite{yeo2001introductory}
&Chinese, English, Japanese, Portuguese&THERM\\
TCS&Thermodynamic Concept Survey&Thermal / Statistical content knowledge (temperature, heat transfer, ideal gas law, first law of thermodynamics, phase change, thermal properties of materials)&\cite{wattanakasiwich2013construction}&2&
43~\cite{wattanakasiwich2013construction}\newline
46~\cite{wattanakasiwich2013construction}\newline
58~\cite{wattanakasiwich2013construction}
&Chinese, English, Thai&THERM\\
TOAST&Test of Astronomy Standards&Astronomy content knowledge (gravity, electromagnetic radiation, fusion and formation of heavy elements, evolution of the universe, star and stellar evolution, evolution and structure of the solar system, seasons, scale, yearly patterns, daily patterns, moon phases)&\cite{slater2014development}&vf&
44~\cite{slater2014development}
&English, Japanese&AST\\

\multirow{2}{*}{TUG-K} &
\multirow{2}{3.4cm}{Test of Understanding Graphs in Kinematics} &
\multirow{2}{5cm}{Mechanics content knowledge (kinematics, graphing)} &
\multirow{2}{*}{\cite{beichner1994testing}} &
2.6 & 
\multirow{2}{*}{59~\cite{klein2020visual}}
&Arabic,  Finnish, French, German, Hebrew  &
\multirow{2}{*}{MECH} \\
&&&&3.0-4.0&&  Chinese, English, Greek,  Portuguese, Spanish, Swedish, Ukrainian & \\

TUV&Test of Understanding of Vectors&Mathematics content knowledge (magnitude, direction, components, unit vector, addition, subtraction, multiplication, dot and cross product)&\cite{barniol2014test}&1&
68~\cite{barniol2014test}
&Arabic, English, Spanish&MATH

\end{tabular}
\end{ruledtabular}
\label{tab:inventories4}
\end{table*}

We cannot exclude the possibility that some of these inventories appeared in the text corpus used to train the model, which could be seen as ``teaching to the test.'' However, GPT's training corpus is closed-source and its contents remain mostly speculative with only very few hints given~\cite{gpttrain}. While popular inventories like the FCI might have been included, it is less likely that many of the more obscure inventories were present. Furthermore, in the scientific literature, the solution keys are typically provided separately from the inventory items. This means that the model is unlikely to associate correct answers to their corresponding questions on the basis of proximity in the training data.

\section{Methodology}
\subsection{Data preparation}
The items from the concept inventories were captured using screenshots like the ones shown in Fig.~\ref{fig:examples}. If an item had multiple parts referring to the same scenario or each other, these were combined in one image; at times, this required manual image-editing to close page breaks. The inventories in Tables~\ref{tab:inventories1}--\ref{tab:inventories4} resulted in 3,662 separate image files that were submitted to the model.

The solution keys were transcribed from PhysPort, where we transformed the various option values such as B), 2., b., $\beta$), and non-Arabic, non-Latin, non-Greek characters into lower-case Latin characters for easier automatic processing.

For the MUQ, which assigns partial credit for incorrect answers, we simplified the evaluation by only giving credit for correct answers; this would under-estimate the model's performance. We did the same for BEMA, which has conditional grading rules for four out of 31~items, depending on answers from earlier items. Here, for two items, this simplification is in the model's favor, while for the other two items, it is to the model's disadvantage. In the FTGOT, we excluded the items assessing certainty in the answer and compared results for a two-tier test version. Finally, we skipped the free-response items that were included in some of the inventories, for example to explain reasoning, as those were not scored in the original inventories.

For each individual item, we coded if it was text-only, or if it included an image, graph, or scenario illustration. Additionally, we coded if consulting the image was necessary for correctly answering the question (required image), or if all the required information was already present in the text (unneeded image). These codes will be used in analyses of subclasses of problems in Section IV.D.

\subsection{AI processing}
We used GPT-4o~\cite{gpt4o} Version 2024-08-06 via Microsoft Azure AI Services~\cite{azure} at ETH~Zurich. The university's contract includes provisions that any data submitted will not be used for training purposes; this provision is crucial to avoid compromising the confidentiality and validity of the concept inventories. 
 
The model was prompted to extract from each submitted image the number and the text of the inventory item, followed by written-out reasoning steps (explanation), and finally the letter option corresponding to its selected answer. In cases where multiple items appeared in the same screenshot, the model was instructed to repeat this process for each item. To process such outputs effectively, we had to prompt the model to return structured outputs in the form of a JSON schema. We found that prompting for structured outputs in languages other than English was unreliable and hindered the type of analysis we aimed to perform. As a result, we decided to keep the prompt in English. Each submitted image, however, contained text in one of the different languages of the concept inventories. The API call and prompts used in this study are available in Appendix~\ref{app:api}.

LLMs are probabilistic systems, and thus responses to the same prompts vary. To obtain some statistics, each screenshot was independently submitted three times, and the resulting three outputs were combined into a solution array.

Altogether, we obtained 14,022 solutions for 4,674 items (1,498 for English and 3,176 for non-English language inventories).

\subsection{Analysis methods}
The LLM's answer choices were normed to the same lower-case Latin characters as the solution keys. In cases where the AI did not provide a valid answer (e.g., where it claimed that there was no correct answer specified), the answer was counted as incorrect. Some concept inventories had images or scenarios labeled with Roman characters and answer choices such as ``A) I, B) II, C) III, D) IV, E) none of the above;'' in this case, if the AI picked ``IV,'' this was manually converted to ``d;'' this happened for 0.4\% of the responses. Another source of possible error were numbered scenarios within numbered multipart problems; in this case, the AI at times ignored the problem numbers provided in the prompt and instead used the scenario numbers; this was fairly obvious during the evaluation, but had to be fixed manually. In 0.6\% of the cases, the LLM provided no valid response; these were counted as incorrect.

We counted each answer in an inventory --- three answers per item --- as either correct or incorrect, and considered the percentage of correct answers across each inventory and language as the performance measure for that test and language. For each answer, we coded the language of both the inventory description and the answer explanation, primarily using \textit{langdetect}~\cite{langdetect} and manual determination in some cases. Responses were categorized as fully in the language of the test, fully in English, or mixed. The most common mixed scenario occurred when the problem description was still in the language of the test, but the explanation was in English. However, there were also rare cases where the language switched mid-stream for either of those. We refer to this behavior as language switching.

\subsection{Use of AI}
While obviously being the subject of this study, AI (GPT-o1~\cite{gpto1}) has also been used for the following aspects of the study: initial drafts of analysis programs in R and Python, exploratory data analysis, LaTeX formatting of manuscript components, and improving the grammar and readability of manuscript passages. 

\section{Results}

\subsection{How does GPT-4o perform across different physics concept inventories in English?}

\begin{figure}
    \centering
    \includegraphics[width=\columnwidth]{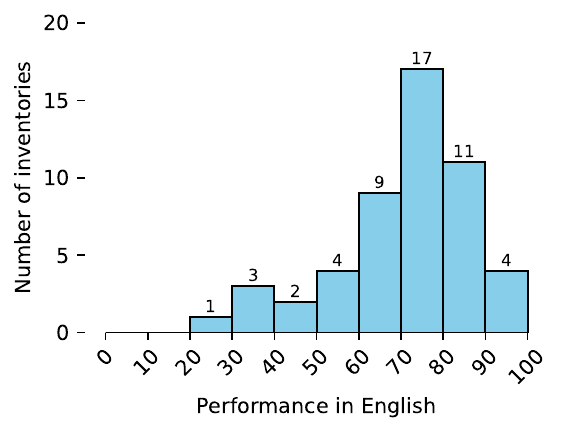}
    \caption{Distribution of scores achieved by GPT-4o on physics concept inventories administered in English.}
    \label{fig:perf_range_eng}
\end{figure}

Notably, English was the language with the most concept inventories: 53 out of 54 tested concept inventories were available in English. It is also the only language that has inventories across all the subject categories. In Table~\ref{tab:scores},  the English results for individual inventories stand out as the most populated column. 
The average performance across all inventories in English is 71.1\%, with results on individual inventories ranging from as low as 27\% (FTGOT) to as high as 97\% (DS and STPFaSLlo). Figure~\ref{fig:perf_range_eng} shows the distribution of GPT-4o's performance across the different concept inventories in English.

\begin{figure}
    \centering
    \includegraphics[width=\columnwidth]{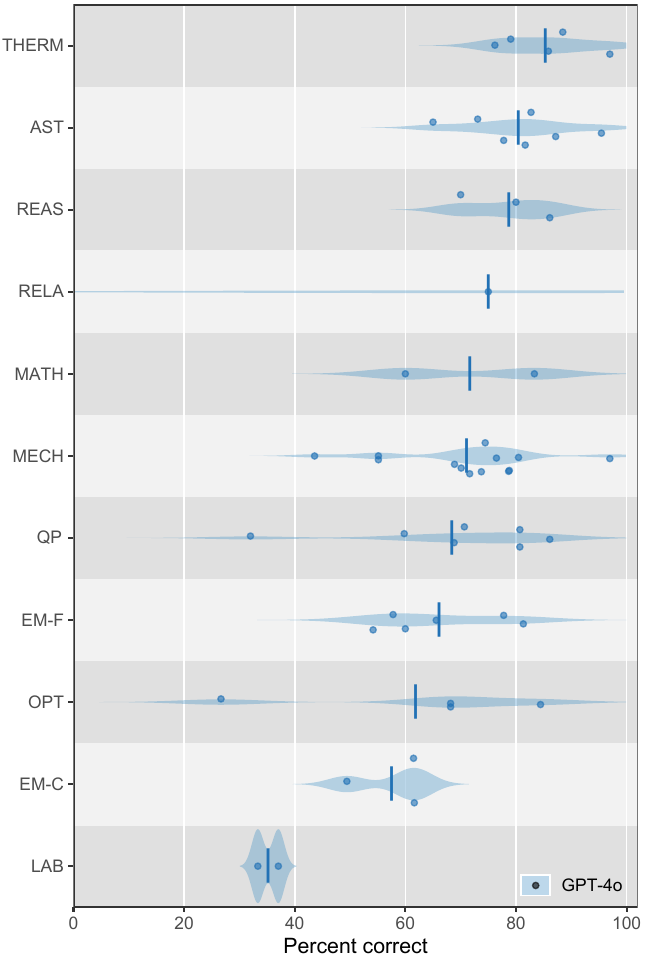}
    \caption{Sina plot~\cite{sidiropoulos2018sinaplot} of GPT-4o's performance on English-language concept inventories, grouped by category (see Table~\ref{tab:sub_cat_used}). The categories are sorted by average performance, indicated by the vertical markers.}
    \label{fig:avg_by_cat}
\end{figure}

Examining the performance within specific subject categories, it is noticeable that for some subjects, the performance varies widely across inventories. As Fig.~\ref{fig:avg_by_cat} shows, this includes outliers in the 30\% range (that is, only slightly better than randomly picking answer options):
\begin{itemize}
\item Within quantum physics (QP), this outlier is the Quantum Mechanics Visualization Inventory (QMVI) at 32\%. This inventory heavily focuses in graphical visualizations of wave functions.
\item Within optics (OPT), the outlier is the Four-tier Geometrical
Optics Test (FTGOT) at 26\%. This inventory deals with ray optics, another graphical visualization topic.
\end{itemize}
However, the extreme values on individual inventories mostly average out --- except in categories with few inventories (e.g., Relativity and Laboratory skills).

The AI performs best in Thermodynamics (85.2\%), followed by Astronomy (80.4\%), Reasoning (78.7\%), and Relativity (75.0\%). Inventories in Astronomy tend to focus on factual knowledge, while those in Thermodynamics, Reasoning, and Relativity tend to use exact language. The AI's performance is weaker in Laboratory skills (35.0\%), which include strategies for data collection and analysis.

\subsection{How does language influence the performance of the AI system?}

Table~\ref{tab:scores} also shows the scores on each inventory in its respective nominal language (we use ``nominal'' to denote the language in which the inventory was presented). One immediate observation is the uneven coverage of assessments across languages. Many inventories are available in only a handful of languages, making broad comparisons challenging. Some assessments, however --- most notably FCI --- are available in numerous languages.

In the FCI, performance ranges from as low as 20\% in Punjabi and 22\% in Tamil to as high as 74\% in Portuguese and Polish, suggesting different performance across languages. Since the FCI uses five answer options, a 20\% performance is equivalent to random guessing.

Similarly, MBT scores fluctuate widely --- e.g., 27\% in Persian, 53\% in Finnish, and 44\% in Italian and Hungarian (SD=$\pm6\%$; Range=$26\%$) --- with Finnish performance nearly double that of Persian.
In contrast, some inventories show more consistent performance across multiple languages. For example, QMCS shows high and relatively stable scores (78–86\%) in the available languages (SD=$\pm4\%$; Range=$8\%$), indicating that the AI handles this conceptual domain and its translations fairly well. TUV, ADT, and FMCE, also show relatively stable performance in the 60–80\% range. Therefore, these results suggest that performance stability across languages is inventory-dependent.

\begin{figure}
\begin{center}
\includegraphics[width=\columnwidth]{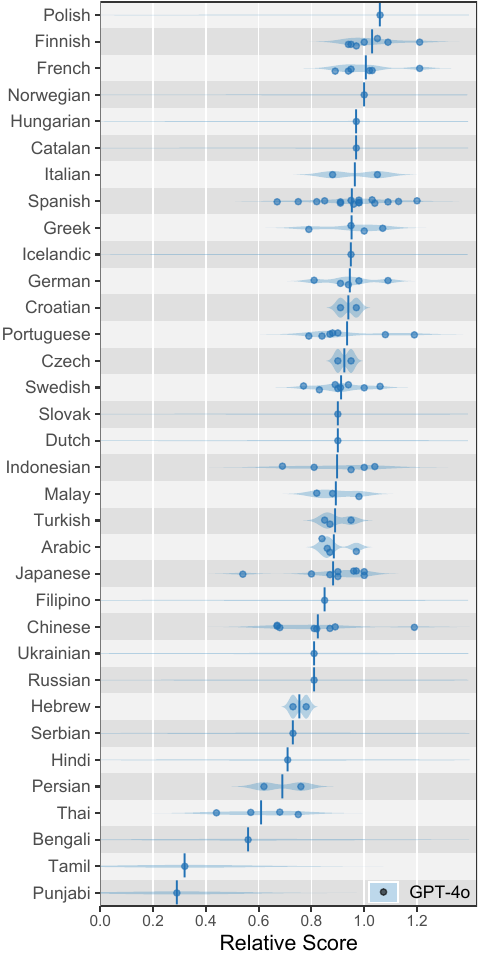}
\end{center}
\caption{Sina plots~\cite{sidiropoulos2018sinaplot} of GPT-4o's performance on inventories in different nominal languages, relative to performance on the same tests in English. The English performance on each inventory was normed as unity, and the plots show the distribution of other-language relative performance. The nominal languages are sorted by average relative performance, indicated by the vertical markers.}
\label{fig:languages}
\end{figure}

Out of 36 inventories available in both English and at least one other language, English was the best-performing language in 27 cases (75\%). Figure~\ref{fig:languages} shows the relative performance by language compared to English. For some languages, only a single inventory was available --- usually the FCI. Notably for Finnish and French, where we have several translated inventories available, the performance is similar to that in English, with a number of other languages, mostly European, trailing closely behind. On the other hand, performance is much lower in languages such as Persian and Thai.

Merely examining performance on an inventory in a given language is insufficient to determine whether the AI struggles with the same items across languages. Although a detailed item-level analysis is beyond the scope of this manuscript, we provide a brief investigation to assess whether items that the AI finds challenging in English remain so in their non-English versions. One indicator of item difficulty is the agreement among the three independent responses and how consistently the AI system provides the correct answers. Since all inventories --- except TUG-K2.6 --- were presented to the AI in English, we can directly compare GPT-4o's performance on English items to their translated counterparts. Our findings indicate that items deemed difficult in English tend to remain challenging in other languages, and vice versa. When all three English responses were incorrect, the correct response rate for the non-English versions was only 18\%. This increased to 33\% when one English response was correct, 48\% for two, and 82\% for three.  Table~\ref{tab:score_quest} in Appendix~\ref{app:Question} illustrates that this trend --- where the same items are consistently challenging across various languages --- holds true for most languages in our dataset.

An interesting finding is that GPT-4o often exhibits language-switching behavior with non-English inventory items. Portuguese and Spanish were the only two languages where the majority of answers (56\% and 59\%, respectively) were entirely in the nominal language. In all other cases, the model predominantly switched into English, either fully or in part.  ~Table \ref{tab:perf_lang} in Appendix~\ref{app:Question} provides additional details on this language-switching behavior.

\begin{table*}
\caption{Scores in percent on each of the inventories in each of the available languages. Green indicates that the AI-score is higher than the average student post-test scores found in literature for
undergraduate-level courses (see column ``\%Post" in Tables~\ref{tab:inventories1}-\ref{tab:inventories4}); red indicates lower AI-performance. A blue background indicates that no student score was available.}

\begin{tabular}{lccccccccccccccccccccccccccccccccccc}
\hline\hline\\
 & \rotatebox{-90}{Arabic} & \rotatebox{-90}{Bengali} & \rotatebox{-90}{Catalan} & \rotatebox{-90}{Chinese} & \rotatebox{-90}{Croatian} & \rotatebox{-90}{Czech} & \rotatebox{-90}{Dutch} & \rotatebox{-90}{English} & \rotatebox{-90}{Filipino} & \rotatebox{-90}{Finnish} & \rotatebox{-90}{French} & \rotatebox{-90}{German} & \rotatebox{-90}{Greek} & \rotatebox{-90}{Hebrew} & \rotatebox{-90}{Hindi} & \rotatebox{-90}{Hungarian} & \rotatebox{-90}{Icelandic} & \rotatebox{-90}{Indonesian} & \rotatebox{-90}{Italian} & \rotatebox{-90}{Japanese} & \rotatebox{-90}{Malay} & \rotatebox{-90}{Norwegian} & \rotatebox{-90}{Persian} & \rotatebox{-90}{Polish} & \rotatebox{-90}{Portuguese} & \rotatebox{-90}{Punjabi} & \rotatebox{-90}{Russian} & \rotatebox{-90}{Serbian} & \rotatebox{-90}{Slovak} & \rotatebox{-90}{Spanish} & \rotatebox{-90}{Swedish} & \rotatebox{-90}{Tamil} & \rotatebox{-90}{Thai} & \rotatebox{-90}{Turkish} & \rotatebox{-90}{Ukrainian} \\
\hline

ADT & - & - & - & - & - & - & - & \cellcolor{GreenYellow}{78} & - & - & - & - & - & - & - & - & - & - & - & - & - & - & - & - & - & - & - & - & - & \cellcolor{GreenYellow}{81} & \cellcolor{GreenYellow}{83} & - & - & - & - \\ 
BEMA & - & - & - & \cellcolor{Apricot}{44} & - & - & - & \cellcolor{GreenYellow}{66} & - & - & - & - & - & - & - & - & - & - & - & \cellcolor{GreenYellow}{53} & - & - & - & - & \cellcolor{GreenYellow}{57} & - & - & - & - & \cellcolor{GreenYellow}{49} & \cellcolor{GreenYellow}{51} & - & - & - & - \\ 
CCI & - & - & - & - & - & \cellcolor{GreenYellow}{79} & - & \cellcolor{GreenYellow}{83} & - & - & - & - & - & - & - & - & - & - & - & - & - & - & - & - & - & - & - & - & - & - & - & - & - & - & - \\ 
CDPA & - & - & - & - & - & - & - & \cellcolor{Apricot}{33} & - & - & - & - & - & - & - & - & - & - & - & - & - & - & - & - & - & - & - & - & - & \cellcolor{GreenYellow}{40} & - & - & - & - & - \\ 
CSEM & - & - & - & - & - & - & - & \cellcolor{Apricot}{54} & - & - & - & - & - & - & - & - & - & \cellcolor{Apricot}{38} & - & - & \cellcolor{Apricot}{53} & - & - & - & - & - & - & - & - & \cellcolor{Apricot}{53} & \cellcolor{Apricot}{45} & - & - & - & - \\ 
CTSR & - & - & - & - & - & - & - & \cellcolor{GreenYellow}{86} & - & - & - & - & - & - & - & - & - & - & - & - & - & - & - & - & - & - & - & \cellcolor{Apricot}{62} & - & \cellcolor{GreenYellow}{74} & \cellcolor{GreenYellow}{76} & - & - & - & - \\ 
DIRECT & - & - & - & \cellcolor{GreenYellow}{59} & - & - & - & \cellcolor{Apricot}{49} & - & \cellcolor{Apricot}{52} & - & \cellcolor{Apricot}{45} & \cellcolor{Apricot}{53} & - & - & - & - & - & - & - & - & - & - & - & - & - & - & - & - & \cellcolor{Apricot}{33} & \cellcolor{Apricot}{49} & - & - & - & - \\ 
DS & - & - & - & - & - & - & - & \cellcolor{GreenYellow}{97} & - & - & - & \cellcolor{GreenYellow}{79} & - & - & - & - & - & - & - & - & - & - & - & - & - & - & - & - & - & - & - & - & - & - & - \\ 
ECA & - & - & - & - & \cellcolor{GreenYellow}{72} & - & - & \cellcolor{GreenYellow}{79} & - & - & - & - & - & - & - & - & - & - & - & - & - & - & - & - & - & - & - & - & - & - & - & - & - & - & - \\ 
ECCE & - & - & - & - & - & - & - & \cellcolor{GreenYellow}{61} & - & - & - & - & - & - & - & - & - & - & - & - & - & - & - & - & - & - & - & - & - & - & - & - & - & - & - \\ 
EMCA & - & - & - & - & - & - & - & \cellcolor{GreenYellow}{78} & - & - & - & - & - & - & - & - & - & \cellcolor{GreenYellow}{78} & - & - & - & - & - & - & - & - & - & - & - & - & - & - & - & - & - \\ 
EMCS & - & - & - & - & - & - & - & \cellcolor{GreenYellow}{79} & - & \cellcolor{GreenYellow}{75} & - & - & - & - & - & - & - & \cellcolor{GreenYellow}{75} & - & - & - & - & - & - & - & - & - & - & - & - & \cellcolor{GreenYellow}{71} & - & - & - & - \\ 
FCI & \cellcolor{GreenYellow}{58} & \cellcolor{Apricot}{39} & \cellcolor{GreenYellow}{67} & \cellcolor{GreenYellow}{57} & \cellcolor{GreenYellow}{67} & \cellcolor{GreenYellow}{62} & \cellcolor{GreenYellow}{62} & \cellcolor{GreenYellow}{69} & \cellcolor{GreenYellow}{59} & \cellcolor{GreenYellow}{67} & \cellcolor{GreenYellow}{70} & \cellcolor{GreenYellow}{68} & \cellcolor{GreenYellow}{66} & \cellcolor{Apricot}{50} & \cellcolor{Apricot}{49} & \cellcolor{GreenYellow}{67} & \cellcolor{GreenYellow}{66} & - & \cellcolor{GreenYellow}{72} & \cellcolor{GreenYellow}{60} & \cellcolor{GreenYellow}{57} & \cellcolor{GreenYellow}{69} & \cellcolor{Apricot}{52} & \cellcolor{GreenYellow}{73} & \cellcolor{GreenYellow}{74} & \cellcolor{Apricot}{20} & \cellcolor{GreenYellow}{56} & - & \cellcolor{GreenYellow}{62} & \cellcolor{GreenYellow}{68} & \cellcolor{GreenYellow}{64} & \cellcolor{Apricot}{22} & \cellcolor{Apricot}{47} & \cellcolor{GreenYellow}{66} & - \\ 
FMCE & - & - & - & - & - & - & - & \cellcolor{GreenYellow}{72} & - & - & - & - & - & - & - & - & - & \cellcolor{GreenYellow}{74} & - & \cellcolor{Apricot}{39} & - & - & - & - & - & - & - & - & - & \cellcolor{GreenYellow}{81} & - & - & - & - & - \\ 
FORT & - & - & - & - & - & - & - & \cellcolor{GreenYellow}{70} & - & - & - & - & - & - & - & - & - & - & - & - & - & - & - & - & - & - & - & - & - & - & - & - & - & - & - \\ 
FTGOT & - & - & - & - & - & - & - & \cellcolor{GreenYellow}{27} & - & - & - & - & - & - & - & - & - & - & - & - & - & - & - & - & - & - & - & - & - & - & - & - & - & \cellcolor{GreenYellow}{23} & - \\ 
FVA & - & - & - & - & - & - & - & \cellcolor{GreenYellow}{76} & - & - & - & - & - & - & - & - & - & - & - & - & - & - & - & - & - & - & - & - & - & - & - & - & - & - & - \\ 
GECI & - & - & - & - & - & - & - & \cellcolor{GreenYellow}{82} & - & - & - & - & - & - & - & - & - & - & - & \cellcolor{GreenYellow}{82} & - & - & - & - & - & - & - & - & - & - & - & - & - & - & - \\ 
HTCE & - & - & - & \cellcolor{Apricot}{68} & - & - & - & \cellcolor{Apricot}{76} & - & - & - & - & - & - & - & - & - & - & - & - & - & - & - & - & - & - & - & - & - & - & - & - & - & - & - \\ 
IBCDC & - & - & - & - & - & - & - & \cellcolor{GreenYellow}{62} & - & - & \cellcolor{GreenYellow}{64} & - & - & - & - & - & - & - & - & - & - & - & - & - & - & - & - & - & - & - & - & - & - & - & - \\ 
IBCM & - & - & - & - & - & - & - & \cellcolor{GreenYellow}{74} & - & - & \cellcolor{GreenYellow}{66} & - & - & - & - & - & - & - & - & - & - & - & - & - & - & - & - & - & - & - & - & - & - & - & - \\ 
LPCA & - & - & - & - & - & - & - & \cellcolor{GreenYellow}{84} & - & - & - & - & - & - & - & - & - & - & - & - & - & - & - & - & - & - & - & - & - & - & - & - & - & - & - \\ 
LPCI & - & - & - & - & - & - & - & \cellcolor{GreenYellow}{65} & - & - & - & - & - & - & - & - & - & - & - & - & - & - & - & - & - & - & - & - & - & \cellcolor{GreenYellow}{62} & - & - & - & - & - \\ 
LSCI & - & - & - & - & - & - & - & \cellcolor{GreenYellow}{73} & - & - & - & - & - & - & - & - & - & - & - & - & - & - & - & - & - & - & - & - & - & - & - & - & - & - & - \\ 
MBT & - & - & - & - & - & - & - & \cellcolor{Apricot}{44} & - & \cellcolor{Apricot}{53} & \cellcolor{Apricot}{41} & \cellcolor{Apricot}{41} & \cellcolor{Apricot}{44} & - & - & - & - & - & \cellcolor{Apricot}{38} & \cellcolor{Apricot}{44} & \cellcolor{Apricot}{38} & - & \cellcolor{Apricot}{27} & - & \cellcolor{Apricot}{35} & - & - & - & - & \cellcolor{Apricot}{40} & - & - & - & \cellcolor{Apricot}{37} & - \\ 
MCS & - & - & - & - & - & - & - & \cellcolor{GreenYellow}{58} & - & - & - & - & - & - & - & - & - & - & - & - & - & - & - & - & - & - & - & - & - & - & - & - & - & - & - \\ 
MUQ & - & - & - & - & - & - & - & \cellcolor{SkyBlue}{37} & - & - & - & - & - & - & - & - & - & - & - & - & - & - & - & - & - & - & - & - & - & - & - & - & - & - & - \\ 
MWCS1 & - & - & - & - & - & - & - & \cellcolor{GreenYellow}{79} & - & - & - & - & - & - & - & - & - & - & - & - & - & - & - & - & - & - & - & - & - & \cellcolor{GreenYellow}{76} & - & - & \cellcolor{Apricot}{35} & - & - \\ 
MWCS2 & - & - & - & - & - & - & - & \cellcolor{GreenYellow}{68} & - & - & - & - & - & - & - & - & - & - & - & - & - & - & - & - & - & - & - & - & - & \cellcolor{GreenYellow}{62} & - & - & - & - & - \\ 
NGCI & \cellcolor{GreenYellow}{76} & - & - & - & - & - & - & \cellcolor{GreenYellow}{87} & - & - & - & - & - & - & - & - & - & - & - & - & - & - & - & - & - & - & - & - & - & - & - & - & - & - & - \\ 
NGPSD & - & - & - & - & - & - & - & \cellcolor{SkyBlue}{95} & - & - & - & - & - & - & - & - & - & - & - & - & - & - & - & - & - & - & - & - & - & - & - & - & - & - & - \\ 
PIQL & - & - & - & - & - & - & - & \cellcolor{GreenYellow}{80} & - & - & - & - & - & - & - & - & - & - & - & - & - & - & - & - & - & - & - & - & - & - & - & - & - & - & - \\ 
QMCA5 & - & - & - & - & - & - & - & \cellcolor{GreenYellow}{74} & - & - & - & - & - & - & - & - & - & - & - & - & - & - & - & - & \cellcolor{GreenYellow}{67} & - & - & - & - & - & - & - & - & - & - \\ 
QMCA6 & - & - & - & - & - & - & - & \cellcolor{GreenYellow}{81} & - & - & - & - & - & - & - & - & - & - & - & - & - & - & - & - & - & - & - & - & - & - & - & - & - & - & - \\ 
QMCS & - & - & - & - & - & - & - & \cellcolor{GreenYellow}{86} & - & \cellcolor{GreenYellow}{81} & - & - & - & - & - & - & - & - & - & \cellcolor{GreenYellow}{78} & - & - & - & - & - & - & - & - & - & - & - & - & - & - & - \\ 
QMFPS & - & - & - & - & - & - & - & \cellcolor{GreenYellow}{60} & - & - & - & - & - & - & - & - & - & - & - & - & - & - & - & - & - & - & - & - & - & \cellcolor{GreenYellow}{49} & - & - & - & - & - \\ 
QMS & - & - & - & - & - & - & - & \cellcolor{GreenYellow}{69} & - & - & - & - & - & - & - & - & - & - & - & - & - & - & - & - & - & - & - & - & - & - & - & - & - & - & - \\ 
QMVI & - & - & - & - & - & - & - & \cellcolor{Apricot}{32} & - & - & - & - & - & - & - & - & - & - & - & - & - & - & - & - & - & - & - & - & - & - & - & - & - & - & - \\ 
QPCS & - & - & - & - & - & - & - & \cellcolor{GreenYellow}{71} & - & - & - & - & - & - & - & - & - & - & - & - & - & - & - & - & - & - & - & - & - & - & - & - & \cellcolor{Apricot}{40} & - & - \\ 
RAPT & - & - & - & - & - & - & - & \cellcolor{GreenYellow}{81} & - & - & - & - & - & - & - & - & - & - & - & - & - & - & - & - & - & - & - & - & - & - & - & - & - & - & - \\ 
RCI & - & - & - & - & - & - & - & \cellcolor{GreenYellow}{75} & - & - & - & - & - & - & - & - & - & - & - & - & - & - & - & - & - & - & - & - & - & - & - & - & - & - & - \\ 
RFCI & - & - & - & - & - & - & - & \cellcolor{GreenYellow}{80} & - & \cellcolor{GreenYellow}{80} & - & - & - & - & - & - & - & - & - & - & - & - & - & - & - & - & - & - & - & - & - & - & - & - & - \\ 
RKI & - & - & - & - & - & - & - & \cellcolor{GreenYellow}{70} & - & - & \cellcolor{GreenYellow}{67} & - & - & - & - & - & - & - & - & - & - & - & - & - & - & - & - & - & - & - & - & - & - & - & - \\ 
RRMCS & - & - & - & - & - & - & - & \cellcolor{Apricot}{74} & - & - & - & - & - & - & - & - & - & - & - & - & - & - & - & - & - & - & - & - & - & - & - & - & - & - & - \\ 
SGCE & - & - & - & - & - & - & - & \cellcolor{GreenYellow}{60} & - & - & - & - & - & - & - & - & - & - & - & - & - & - & - & - & - & - & - & - & - & - & - & - & - & - & - \\ 
SPCI & - & - & - & - & - & - & - & \cellcolor{GreenYellow}{95} & - & - & - & - & - & - & - & - & - & - & - & \cellcolor{GreenYellow}{86} & - & - & - & - & - & - & - & - & - & \cellcolor{GreenYellow}{94} & - & - & - & - & - \\ 
STPFaSLSh & - & - & - & \cellcolor{GreenYellow}{59} & - & - & - & \cellcolor{GreenYellow}{86} & - & - & - & - & - & - & - & - & - & \cellcolor{GreenYellow}{70} & - & - & - & - & - & - & \cellcolor{GreenYellow}{72} & - & - & - & - & - & - & - & - & - & - \\ 
STPFaSLlo & - & - & - & - & - & - & - & \cellcolor{SkyBlue}{97} & - & - & - & - & - & - & - & - & - & - & - & - & - & - & - & - & - & - & - & - & - & - & - & - & - & - & - \\ 
TCE & - & - & - & \cellcolor{Apricot}{59} & - & - & - & \cellcolor{GreenYellow}{88} & - & - & - & - & - & - & - & - & - & - & - & \cellcolor{GreenYellow}{85} & - & - & - & - & \cellcolor{GreenYellow}{78} & - & - & - & - & - & - & - & - & - & - \\ 
TCS & - & - & - & \cellcolor{GreenYellow}{69} & - & - & - & \cellcolor{GreenYellow}{79} & - & - & - & - & - & - & - & - & - & - & - & - & - & - & - & - & - & - & - & - & - & - & - & - & \cellcolor{GreenYellow}{59} & - & - \\ 
TOAST & - & - & - & - & - & - & - & \cellcolor{GreenYellow}{83} & - & - & - & - & - & - & - & - & - & - & - & \cellcolor{GreenYellow}{80} & - & - & - & - & - & - & - & - & - & - & - & - & - & - & - \\ 
TUG-K2.6 & \cellcolor{Apricot}{48} & - & - & - & - & - & - & - & - & \cellcolor{GreenYellow}{60} & \cellcolor{GreenYellow}{67} & \cellcolor{GreenYellow}{60} & - & \cellcolor{Apricot}{43} & - & - & - & - & - & - & - & - & - & - & - & - & - & - & - & - & - & - & - & - & - \\ 
TUG-K3.0-4.0 & - & - & - & \cellcolor{Apricot}{45} & - & - & - & \cellcolor{Apricot}{55} & - & - & - & - & \cellcolor{Apricot}{44} & - & - & - & - & - & - & - & - & - & - & - & \cellcolor{GreenYellow}{65} & - & - & - & - & \cellcolor{GreenYellow}{60} & \cellcolor{Apricot}{50} & - & - & - & \cellcolor{Apricot}{45} \\ 
TUV & \cellcolor{Apricot}{58} & - & - & - & - & - & - & \cellcolor{Apricot}{60} & - & - & - & - & - & - & - & - & - & - & - & - & - & - & - & - & - & - & - & - & - & \cellcolor{Apricot}{62} & - & - & - & - & - \\ 

\hline\hline
\end{tabular}
\label{tab:scores}
\end{table*}

\subsection{How does the AI's performance compare to student performance at the undergraduate level?}

In Table~\ref{tab:scores} we use color to indicate the relative performance of GPT-4o compared to published post-instruction scores at the undergraduate student level found in the literature (see column ``\%Post'' in Tables~\ref{tab:inventories1}-\ref{tab:inventories4}). For most inventories in most languages, the AI system outperforms the student average. Where student data was available, the AI outperformed average undergraduate post-instruction scores in 68.9\% of cases. To obtain this value, we compared student averages on an inventory to AI's performance on the inventory for each available language and summed across all inventories.

As we have pointed out in the Introduction (Section~\ref{sec:relevant}), the comparison of average performance of GPT-4o to that of students provides a rough proxy measure for its capabilities in solving physics conceptual tasks in relation to student capabilities. However, caution is needed when interpreting these results, as the profile of GPT-4o's and the "average" student's strengths and difficulties can differ. 

\begin{figure}
\begin{center}
\includegraphics[width=\columnwidth]{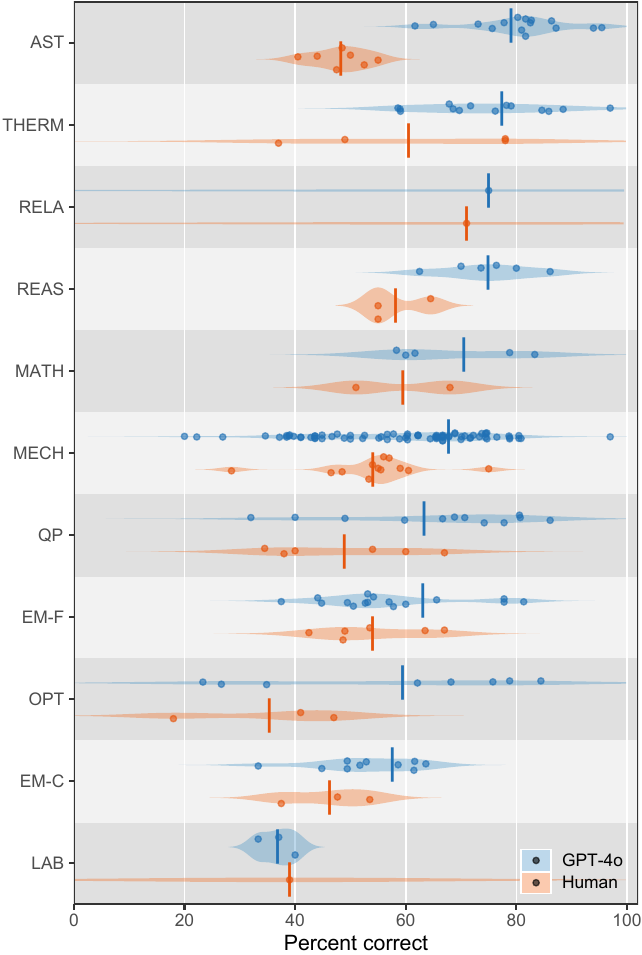}
\end{center}
\caption{Sina plots~\cite{sidiropoulos2018sinaplot} of GPT-4o (all languages) and student scores on physics concept inventories grouped by subject category.
 The categories are arranged in descending order of the average GPT‑4o score; average scores are indicated by vertical markers.}
\label{fig:perf_subj}
\end{figure}

Figure~\ref{fig:perf_subj} shows the distributions of post-test student scores and AI scores across all languages, grouped by subject category. The averages are indicated by vertical markers. With the exception of Laboratory skills (LAB), GPT-4o outperforms the average undergraduate student in every subject category of concept inventories, with the most significant differences in Astronomy and Reasoning.
The extremely wide distribution of the mechanics
(MECH) scores is mostly due to the fact that the Force
Concept Inventory (FCI) is available in a wide variety
of languages, including languages such as Punjabi and
Tamil, which GPT-4o does not appear to adequately
master (see Fig.~\ref{fig:languages}). The wide distributions in Quantum Physics and Optics are again due to inventories with mostly graphical, visual representations.

Well-designed assessment instruments for teaching typically include multiple-choice distractors that probe specific student misconceptions. Since we drew our questions from the PhysPort library, we can assume that most concept inventories in our dataset were designed carefully. Therefore, we expect students to gravitate toward certain incorrect answers that reflect common misconceptions. This raises the question of whether the AI similarly gravitates toward specific incorrect answers. Although not all the assessments analyzed in this study had five possible choices, the majority did, allowing us to use this as a reference point (see Appendix~\ref{app:Question} and Table~\ref{tab:incorrect_Q} for more detail). If the AI were selecting incorrect answers randomly, we would expect a 75\% probability that two incorrect answers differ and a 25\% chance that they would be the same. However, for our data, this was reversed: 66\% of all items had the same two incorrect answers, while 34\% had different ones. The same pattern held for cases where all three answers were incorrect. In a random scenario,we would expect 6.25\% of items to have three identical answers, 37.5\% to have three different ones, and 56.25\% to have two the same (see Appendix~\ref{app:Question} for more details). In reality, AI responses were far from random and gravitate toward specific incorrect answers: 53\%  of such items had three identical incorrect answers, while only 8\% had three different ones. Similar trends were observed for both English and non-English items. This analysis suggests that the AI system exhibits some consistency in its choice of incorrect answers. However, further research is necessary to investigate how students and AI systems may differ in how they respond to different types of multiple-choice options. 

\subsection{How does the presence of images influence the performance of the AI system?
}
\begin{figure}
\begin{center}
\includegraphics[width=\columnwidth]{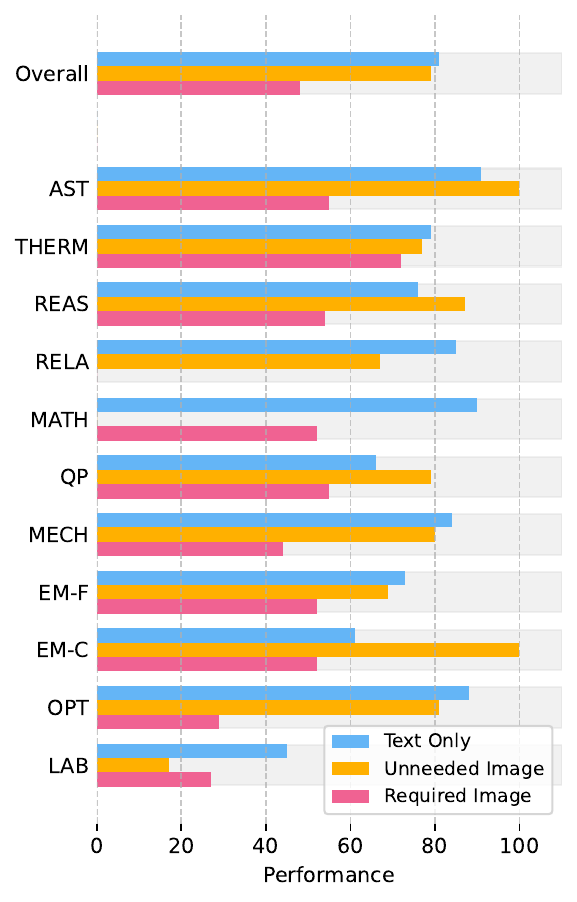}
\end{center}
\caption{Performance on inventory items categorized by their usage of images and grouped by their subject category. Note that in contrast to the inventory-level analysis presented in previous sections, the values on this histogram present the summative performance on all submissions of inventory items belonging to a given image category, collected across the concept inventories in each subject category.}
\label{fig:perf_all_cat}
\end{figure}

Here, we calculate the performance based on the sum across individual inventory items appearing in the different inventories (all languages combined). For each item, we coded if it contained or referred to an image --- that is, a visual representation such as a sketch, graph, diagram --- and whether interpreting the image was required for correctly solving the task (``required image''), or if the image was redundant, meaning all information required for solving the task was already provided in the text (``unneeded image''). The percentages shown in Figure~\ref{fig:perf_all_cat}  were obtained by dividing the number of correct responses in each category with the total number of submissions in that category. 

Overall, we found that the performance on text-only tasks was 81\% , compared to 79\% on tasks containing unneeded images, and just 49\% on tasks with required images (see Figure~\ref{fig:perf_all_cat}).

When examining individual subject categories, GPT-4o consistently performs worse on items that require image interpretation than on text-only items. Relativity, Optics, Mechanics, and Mathematics (which had no unneeded image items), as well as Astronomy, exhibited especially large performance gaps. Furthermore, while QP as a whole was not among the lowest-performing subject categories on image-based problem types, QMVI --- entirely image-based and composed predominantly of required-image items --- was the second worst-performing inventory in English (32\%). In line with previous research on the topic~\cite{polverini24}, these findings suggest that visual interpretation is one of GPT-4o's major weaknesses.

\section{Discussion and future research directions}

Unlike many previous studies where inventories were provided as text-only materials, this study provided the AI with a screenshot of the item as it would appear to a student. Using authentic screenshots rather than isolated text offers a closer approximation of the model's behavior in realistic educational settings, especially in multilingual and multimodal contexts. However, this approach also increases the complexity of the input and the analytical demand for interpreting the model outputs. 

Across the majority of inventories and subject categories (with Laboratory skill being the exception), GPT-4o achieves higher scores and outperforms the average reported student post-instruction undergraduate student. It needs to be emphasized that the instruments are research-based and thus carefully designed with regard to psychometric properties, which is reflected in their student score distributions: (i) The inventory items aim for medium difficulty, resulting in average student scores around 50\%, and (ii) The items are designed for high discrimination, leading to broad score distributions. Thus, the finding that the AI performs better than the average student on these inventories broadly means that it performs above 50\% on these instruments. It also means that in almost all cases, some students in the dataset still outperformed GPT-4o. Furthermore, in studies where results for graduate or post-doctoral populations were available, these outperformed GPT-4o.

It is important to note that even when GPT-4o outperformed the average student, this does not mean that it necessarily exhibited a strength and difficulty profile similar to that of a well-performing student. Our findings support previous research (e.g., ~\cite{polverini2024performance}) and suggest that image interpretation is one common GPT-4o difficulty that is not typically seen in students.

This highlights the importance of carefully considering the role of AI in physics assessment: while it can match or exceed undergraduate averages on concept inventories, its performance should not be mistaken for human expert-like reasoning or deep conceptual understanding. For example, a test on which AI performs poorly might not necessarily be difficult for students, and vice versa. Much depends on the type of tasks involved and how they align with the AI's and students' strength and difficulty profiles.

If AI systems like GPT-4o will play a role in the pipeline of assessment design and validation, the role will likely be different from that of a student or a domain expert.  While there exists some research on making an LLM behave like a student with a certain difficulty profile~\cite{kieser23}, more work is needed to establish the feasibility of such use at scale and across different subject areas. On the other hand, because of the increasingly important role AI systems will likely play in physics education, it is worthwhile considering developing assessments specifically for those AI systems that are going to be used in educational contexts. Such assessment would provide curriculum developers and instructors with even more pertinent information about these systems' capabilities and potential in educational contexts. Exactly what such assessment would look like remains unclear. However, it is reasonable to expect that it will be developed in response to the practical needs and demands of the new educational landscape.

On the linguistic side, our results reveal a complex relationship between performance and language. The overall pattern that emerged is that GPT-4o performed better in most cases when tasks were given in English. While some other languages --- mostly European and Latin-script --- exhibited roughly similar performance, others show significantly lower results. One contributing factor for this difference could be that GPT-4o had to handle English language prompts with screenshots using other language scripts, a task that is presumably harder than just dealing with one language or script. Understanding this better would need to be explored further in a future study.  This is particularly importantnt as it has implications for accessibility and utility of AI tools for non-Western language-speaking populations and may risk perpetuating or even exacerbating disparities in access to educational and technological resources worldwide.

Our results also indicate that tasks found to be challenging in English tend to remain so when posed in other languages. Notably, the model often produces consistent patterns of incorrect responses, even though its training data likely differs across languages. This observation raises questions about whether these errors stem from biases in the training data or reflect deeper limitations in the model’s inference process. While our study does not resolve these issues, it highlights that certain items consistently challenge the model across languages. These findings may inform future research on assessment designs that account for the limitations of multimodal AI systems, whether the items are visual or text-based.

Examining the presence and function of visual representations in inventory items reveals a clear pattern. The AI performed better on text-only items and items with redundant visual representations, compared to items where visual interpretation of images was necessary for solving the task correctly. This corroborates previous findings on the topic, suggesting that the AI's visual reasoning remains a major weakness, hampering its ability to engage effectively with graphical or picture formats common in physics tasks~\cite{polverini2024evaluating,polverini2024performance}. Anecdotal evidence in recent months suggests that, compared to GPT-4o, GPT-o3 and o4-mini~\cite{gpto4mini} perform better on physics tasks, including numerical ones, but often still struggle with the interpretation of figures. 

There may be other variables, beyond subject area, language and image presence, which influence the performance of GPT-4o and other AI systems on physics tasks. The mathematical complexity of a task and the presence of redundant information are just two of the many possible aspects that could influence AI's ability to solve a task. More research is needed to develop a better sense of what makes a physics task difficult for an AI system to solve.

Based on our findings, we suggest that future efforts to improve AI performance on physics tasks should focus on enhancing the multimodal processing components and achieving more balanced performance across languages and conceptual domains.

\section{Implications for instruction}

Our study is decidedly exploratory and does not aim to provide direct advice to instructors on how to implement GPT-4o in education. Still, there are some broad findings that emerged from our exploration of GPT-4o's performance on physics concept inventories, which can be informative for physics instructors and curricular developers.

We expect that in the future, multimodal AI systems like GPT-4o will continue to influence the space of physics education through at least two mechanisms. First, their accessibility and availability and their relatively high capabilities make them attractive for learners. They will likely continue to be used by students to help them with physics tasks. Second, AI tools are likely to see increased uptake by educational institutions to support students in their learning as well as to support instructors in their teaching and administrative tasks.

It is important to note once more that while GPT-4o, on the surface, exhibits performance that is numerically better than university students' post-instruction average, a closer look reveals important caveats with this simple interpretation. If students want to use GPT-4o productively and responsibly, they should be aware of its limitations. We believe that instructors should inform students of these drawbacks to mitigate the risks of over-reliance on AI tools, and foster a critical perspective on the outputs these tools generate. This should arguably become one of the newly emerging instructional goals because such skills will remain useful even as students leave education and enter the workforce. As AI becomes part of our everyday and work, evaluating its output becomes an important skill, which cannot be learned by always outsourcing physics reasoning to an AI.
Exposing AI's drawbacks to students can thus also serve as a motivation for students to engage more deeply in learning physics. Our research can help inform such efforts. For example, we have shown that GPT-4o's performance is not equally good across all subject areas of physics and that it often struggles when prompted in non-Western languages. Furthermore, a major drawback is its ability to interpret images.

However, because of the incredible pace of AI progress, sooner or later, physics educators will have to contend with the question of what remains meaningful to teach, when AI systems perform well on many tasks that were previously squarely in the domain of human physics experts. The physics education community will likely need to seriously reflect on whether physics curricula, which have in many cases remained nearly unchanged for decades, should evolve to better reflect the new reality. If we conclude that the kind of conceptual understanding that is being tested by research-based concept inventories is still valuable and important for our students, then these assessments will likely continue to play an important role in evaluating whether our students have reached the desired learning objectives. In such a case, they should arguably be administered so that students cannot use AI to help them. 

In making these important decisions, the physics education community will likely also need to address the following (and other) questions on a continuous basis:
What are the foundational skills that we should not routinely outsource to AI? What are the central tasks that students first need to master themselves, so that they can later judiciously and responsibly outsource them to AI? How do we ensure that access to top-performing AI does not generate or worsen existing divides based on students' economic or ethnic background?

\section{Limitations}
This study is decidedly exploratory and empirical. The preparation of item images was done manually, and some manual cleanup of the data was required. Given more than 3,600 images and the random oddities occurring in over 14,000 solutions generated by a probabilistic system, clerical errors cannot be excluded. Additionally, each screenshot of inventory items was iterated only three times: given the stochastic nature of LLM outputs, this introduces variability that limits the results' generalizability. The study also did not consider the quality of the concept inventories' translations, and lower scores may be due to incorrect or confusing translations.
AI's performance might be dependent on prompts, and the ones used for our study (shown in Figs.~\ref{fig:role} and~\ref{fig:prompt} of Appendix~\ref{app:api}) may not be the best choices. Future studies may look into whether and how prompt-engineering techniques might improve (or worsen) AI's performance.  

Furthermore, it is unclear to what extent the use of an English prompt, combined with inventory text in the nominal language of each inventory, influenced performance across languages and contributed to language-switching behavior; Appendix~\ref{app:switch} discusses some of the preliminary observations. Future research may explore this further by varying the prompting approach, for example, by prompting the system entirely in the nominal language. However, due to the unreliability of structured outputs with non-English prompts (see Appendix~\ref{app:api}), such approach would likely require alternative output formats, more human involvement in the coding of the answers, as well as accurate translations of the prompts into multiple languages.

The model's visual encoder transforms the pixel data into a latent representation that encapsulates both textual and graphical information. This has the side effect that there is no standalone raw OCR (optical character recognition) output that we could compare to the original text in the various Latin and non-Latin scripts used in the study. One possible explanation for the difference in performance between languages could be incorrect recognition of non-Latin characters.

Furthermore, it should be noted that we did not score the correctness of the physics reasoning in the AI-written explanations. Future studies could explore this in more detail, potentially using the data collected in this study~\cite{our_data} (as an example, see Fig.~\ref{fig:output} in Appendix~\ref{app:api}) to evaluate the model's reasoning separately from its final answers. An example of how this could be done can be found in~\cite{polverini24}.

A possible stumbling block for the AI can be the processing of graphical information unrelated to a physics concept or language. For example, for item~7 of the FCI, the multiple-choice options are embedded in a graphic showing a ball swinging in a circular path and are not listed separately in the text. Although the AI frequently gave physically reasonable explanations, across 32 languages and 96 total answers, it selected the correct embedded option only once. For no obvious reason, it chose instead the same incorrect answer an eyebrow-raising 88 times. Once again, this finding aligns with previous studies showing that the graphical layout of images and the spatial arrangement of answer options can play an important role in the AI's selection of answers~\cite{polverini2024performance}.

The performance of GPT-4o, as measured in our study represents a momentary snapshot of one model's capabilities in early 2025. It is very likely that future models will outperform GPT-4o. However, to keep track of such developments, more studies similar to ours will be needed.

Finally, human post-instruction scores were gathered on a best-effort basis, which may introduce additional variability into the comparisons. Moreover, most of the human data came from English-speaking students taking the English versions of the inventories.

Against the background of these limitations, it is clear that our study only scratches the surface of this exciting area of research and there are many open questions that invite further exploration.

\section{Conclusion}

The results of this study underscore the complexity and variability inherent in using multimodal large language models for physics assessment tasks across multiple languages, subject categories, and formats. 

The marked differences in performance across languages highlight that GPT-4o is not equally competent in all tongues. Anecdotally, this is true for many LLMs. This suggests a risk of generating new, as well as maintaining or exacerbating existing inequities in the access to educational resources and technologies across the world.

Based on published student scores on the tested inventories, we found that GPT-4o outperforms undergraduate student post-instruction averages on most inventories, and in all subject categories except laboratory skills. The reasons for the subject's dependence on its performance are not entirely clear. Possible explanations include different levels of representation in the training data and its varying quality across subjects, or potential differences in the inherent difficulty of tasks in the assessments covering different categories.

We have also found that the presence of non-redundant visual representations negatively influences AI's performance across all subject categories. This suggests that the AI's vision abilities still present a major weakness and consequently limit its utility for some educational uses.

In sum, this exploratory study demonstrates that the studied AI system exhibits significant variations in performance depending on the language, conceptual domain, and presence of visual information. The work points toward the need for future improvements in training data diversity, model fine-tuning, and prompt engineering to enhance its performance. It also highlights the need for careful consideration when implementing such AI systems in educational contexts, ensuring that their use is both equitable and aligned with pedagogical goals.

\section*{Data Availability}
Data will be made available on PhysPort for verified community members~\cite{our_data}.
\begin{acknowledgments}
We would like to thank Sam McKagan for all of her support.
\end{acknowledgments}

\bibliography{crunch}

\newpage

\clearpage

\appendix

\section{Implementation of the LLM-API calls}\label{app:api}

The screenshots were submitted via the deployment's API in base64-encoding as shown in Fig.~\ref{fig:call}, setting a temperature of 0.7 (which is the default in the chat clients); newer  reasoning models like GPT-o1 or GPT-o3-mini do not accept temperature  anymore. As role, we used the text shown in Fig.~\ref{fig:role}, and as prompt the text shown in Fig.~\ref{fig:prompt}. Role (job description) and prompt (task description) tend to be slightly redundant; newer models do neither expect nor accept the role parameter anymore.

To facilitate further processing and evaluation of the answers, we provided the structured JSON output schema shown in Fig.~\ref{fig:structure}; this forces the model to provide its output with the given data structure instead of in narrative form. We included the ``problem\_description'' and ``explanation" fields in the output structure to force the AI system to describe the problem ``in its own words'' and to provide reasoning for the response, thereby triggering Chain-of-Thought (CoT)~\cite{polverini2024ejp}.

As LLMs are probabilistic systems, each screenshot was evaluated three times; this resulted in output like the one shown in Fig.~\ref{fig:output}. Each run is independent, and the model may provide different interpretations of the problem, different reasoning steps, and potentially different final answers every time. The firmer the required concept for a problem is anchored in the model's parameters, the less its answers will vary.

\begin{figure}
\begin{lstlisting}
    responseGPT = client.chat.completions.create(
        model="EthelOmni",
        temperature=0.7,
        messages=[
            {"role": "system", "content": role},
            {
                "role": "user",
                "content": [
                    {"type": "text", "text": prompt},
                    {
                        "type": "image_url",
                        "image_url": {
                            "url": image_data
                        }
                    }
                ]
            }
        ],
        max_tokens=1000,
        response_format={
            "type": "json_schema",
            "json_schema": {
                "name": "problem_response",
                "strict": True,
                "schema": json_schema
            }
        }
    )
\end{lstlisting}

\caption{The API call used in this study. ``EthelOmni'' is a GPT-4o Azure deployment.}
\label{fig:call}
\end{figure}

\begin{figure}
\noindent\fbox{
\begin{minipage}{\columnwidth}
\begin{flushleft}
\small
You are a physics and mathematics expert.

You are given images of multiple-choice test questions, which you will answer correctly in JSON format.

You work very carefully, and you strongly favor correct answers over rapid responses.

You do mathematical calculations and derivations step-by-step.

If there are graphics, you consult them more than once if needed in order to get information for your reasoning.
\end{flushleft}
\end{minipage}}
\caption{The role used for this study.}
\label{fig:role}
\end{figure}

\begin{figure}
\noindent\fbox{
\begin{minipage}{\columnwidth}
\begin{flushleft}
\small
You are tasked with correctly solving the problem or problems from the image.

The image should contain problem(s) with the problem identifiers: [Problems]

For this multiple-choice problem or these multiple-choice problems, carefully consider the situation and document your reasoning,
and only then pick the answer choice that aligns with your reasoning.
\end{flushleft}
\end{minipage}}
\caption{The prompt used for this study. Before submission, [Problems] is replaced by the problem numbers contained in the image, e.g., ``3, 4, 5.'' The image is submitted alongside in base64-encoding.}
\label{fig:prompt}
\end{figure}

\begin{figure*}
\begin{lstlisting}
# Define the JSON schema for the structured response
    json_schema = {
        "type": "object",
        "properties": {
            "problems": {
                "type": "array",
                "items": {
                    "type": "object",
                    "properties": {
                        "problem_description": {"type": "string"},
                        "questions": {
                            "type": "array",
                            "items": {
                                "type": "object",
                                "properties": {
                                    "question_number": {"type": "integer"},
                                    "explanation": {"type": "string"},
                                    "correct_answer": {"type": "string"}
                                },
                                "required": ["question_number", "explanation", "correct_answer"],
                                "additionalProperties": False
                            }
                        }
                    },
                    "required": ["problem_description", "questions"],
                    "additionalProperties": False
                }
            }
        },
        "required": ["problems"],
        "additionalProperties": False
    }
\end{lstlisting}

\caption{The data structure used for this study.}
\label{fig:structure}
\end{figure*}

\begin{figure*}
\begin{lstlisting}
{
    "solutions": [
        {
            "problems": [
                {
                    "problem_description": "What happens to a positive charge that is placed at rest in a uniform magnetic field? (A uniform field is one whose strength and direction are the same at all points.)",
                    "questions": [
                        {
                            "question_number": 21,
                            "explanation": "The force on a charge moving in a magnetic field is given by the Lorentz force, F = q(v x B), where q is the charge, v is the velocity, and B is the magnetic field. If the charge is at rest, then v = 0, and thus F = 0. Therefore, there is no force acting on the charge, and it will remain at rest.",
                            "correct_answer": "e"
                        }
                    ]
                }
            ]
        },
        {
            "problems": [
                {
                    "problem_description": "What happens to a positive charge that is placed at rest in a uniform magnetic field? (A uniform field is one whose strength and direction are the same at all points.)",
                    "questions": [
                        {
                            "question_number": 21,
                            "explanation": "A magnetic field exerts a force on a moving charge, given by the Lorentz force equation: F = q(v \u00d7 B), where q is the charge, v is the velocity, and B is the magnetic field. If the charge is at rest, v = 0, and thus the force F = 0. Therefore, there is no force acting on the charge to cause any acceleration or motion. The charge remains at rest as there is no initial velocity and no force to change its state.",
                            "correct_answer": "e"
                        }
                    ]
                }
            ]
        },
        {
            "problems": [
                {
                    "problem_description": "What happens to a positive charge that is placed at rest in a uniform magnetic field? (A uniform field is one whose strength and direction are the same at all points.)",
                    "questions": [
                        {
                            "question_number": 21,
                            "explanation": "A magnetic field exerts a force on a moving charge according to the Lorentz force law, which states that the force is proportional to the velocity of the charge and the magnetic field. The force is given by \\( \\mathbf{F} = q(\\mathbf{v} \\times \\mathbf{B}) \\). If the charge is initially at rest, then the velocity \\( \\mathbf{v} \\) is zero, resulting in zero force. Therefore, a charge placed at rest in a uniform magnetic field will experience no force and will remain at rest.",
                            "correct_answer": "e"
                        }
                    ]
                }
            ]
        }
    ]
}
\end{lstlisting}
\caption{Typical output; each problem is independently solved three times (three ``problems''-blocks inside of ``solutions'').}
\label{fig:output}
\end{figure*}

As Fig.~\ref{fig:output} illustrates, the role and prompt we used can result in mixed-language outputs. While this issue could probably have been avoided by translating the role and prompts into the language of the inventories, doing so for languages using non-ASCII characters would likely have made the mapping between the prompt and the fields in the JSON structure unreliable. Colloquially speaking, ``the AI was allowed to think in the language of its choice.'' Some observations on this behavior are discussed in Appendix~\ref{app:switch}

\section{Language-switching behavior}\label{app:switch}

While the model transforms input into abstract latent representations rather than processing it in any particular human language, the language of the output may provide some insights into these underlying representations. Table~\ref{tab:perf_lang} shows that, for certain languages, overall accuracy increases when responses are generated entirely in English or as a mix with English. For example, for Bengali, Persian, and Punjabi, accuracy is noticeably higher when English is used in the output. This pattern may reflect differences in the effectiveness of the model’s representations for these languages compared to English. In contrast, for languages such as English, Spanish, and French, high correctness is maintained even when responses remain solely in the original language.

Furthermore, when performance on a given item in English is treated as a baseline for difficulty, Table~\ref{tab:switch_diff} indicates that for non-English inputs the output remains in the original language in 24\% of cases that are easier in English (i.e., those with 3/3 correct answers). This proportion is higher than the 15–17\% observed for items where at least one response in English was incorrect. These observations suggest a correlation between the response language and the relative difficulty as measured by performance in English.

\begin{table}
\caption{Language switching of a non-English item based on level of difficulty for the same item in English}
\begin{ruledtabular}
\begin{tabular}{lcccc}

&\multicolumn{4}{c}{Answers correct in English}\\
Language switch  & 0/3 & 1/3 & 2/3 & 3/3 \\
\hline
Switch to English & 69\% & 69\% & 69\% & 63\% \\
Mixed Language & 14\% & 17\% & 15\% & 12\% \\
Nominal Language & 17\% & 15\% & 17\% & 24\% \\
\end{tabular}
\end{ruledtabular}
\label{tab:switch_diff}
\end{table}

\begin{table*}
\caption{Performance by language and language switching in percent}
\begin{ruledtabular}
\begin{tabular}{lccccccc}
&&\multicolumn{2}{c}{Nominal language}&\multicolumn{2}{c}{Switch to English}&\multicolumn{2}{c}{Mixed languages}\\
Language & \%Correct & \% & \%Correct & \% & \%Correct & \% & \%Correct \\
\hline
Arabic &  60 & 6 & 89 & 92 & 58 & 2 & 83 \\
Bengali &  39 & 2 & 100 & 98 & 38 & 0 &  \\
Catalan &  67 & 19 & 94 & 60 & 57 & 21 & 68 \\
Chinese &  58 & 2 & 57 & 97 & 58 & 1 & 20 \\
Croatian &  69 & 11 & 81 & 64 & 69 & 25 & 66 \\
Czech &  69 & 3 & 100 & 58 & 54 & 40 & 89 \\
Dutch &  62 & 34 & 68 & 39 & 46 & 27 & 79 \\
English &  72 & 100 & 72 & - & - & - & - \\
Finnish &  66 & 4 & 86 & 60 & 65 & 36 & 64 \\
Filipino &  59 & 0 &  & 100 & 59 & 0 &  \\
French &  63 & 37 & 74 & 43 & 55 & 20 & 60 \\
German &  56 & 19 & 70 & 40 & 49 & 41 & 56 \\
Greek &  52 & 8 & 63 & 80 & 50 & 11 & 61 \\
Hebrew &  47 & 1 & 100 & 97 & 46 & 2 & 67 \\
Hindi &  49 & 9 & 75 & 90 & 47 & 1 & 0 \\
Hungarian &  67 & 9 & 100 & 70 & 65 & 21 & 58 \\
Icelandic &  66 & 0 &  & 100 & 66 & 0 &  \\
Indonesian &  67 & 26 & 71 & 59 & 67 & 15 & 60 \\
Italian &  57 & 37 & 71 & 53 & 48 & 10 & 47 \\
Japanese &  63 & 8 & 87 & 83 & 60 & 8 & 72 \\
Malay &  50 & 0 &  & 100 & 50 & 0 &  \\
Norwegian &  69 & 13 & 75 & 71 & 62 & 16 & 93 \\
Persian &  40 & 1 & 0 & 98 & 40 & 1 & 100 \\
Polish &  73 & 0 &  & 100 & 73 & 0 &  \\
Portuguese &  64 & 56 & 73 & 34 & 50 & 10 & 63 \\
Punjabi &  20 & 2 & 100 & 93 & 19 & 4 & 0 \\
Russian &  56 & 8 & 57 & 84 & 55 & 8 & 57 \\
Serbian &  62 & 0 &  & 100 & 62 & 0 &  \\
Slovak &  62 & 0 &  & 76 & 62 & 24 & 64 \\
Spanish &  62 & 59 & 63 & 33 & 62 & 8 & 52 \\
Swedish &  60 & 13 & 65 & 56 & 59 & 31 & 58 \\
Tamil &  22 & 1 & 100 & 87 & 19 & 12 & 36 \\
Thai &  47 & 7 & 67 & 86 & 44 & 7 & 55 \\
Turkish &  40 & 18 & 51 & 64 & 39 & 18 & 35 \\
Ukrainian &  45 & 3 & 0 & 79 & 45 & 18 & 50 \\

Overall &  63 & 46 & 71 & 45 & 55 & 9 & 60 \\

\end{tabular}
\end{ruledtabular}
\label{tab:perf_lang}
\end{table*}

Although GPT-4o does not engage in extended Chain-of-Thought reasoning similar to GPT-o1, DeepSeek, or GPT-o3-mini, its selection of output language --- whether remaining in the nominal language, switching entirely to English, or producing a mixed-language response --- emerges from patterns learned during training rather than from explicit instructions. The prompt was provided in English, while the item content was often in another language and accompanied by multimodal elements such as diagrams and sketches. Consequently, the variation in output language and the resulting solution accuracy offer insights into the model’s internal processing of diverse input formats without implying deliberate reasoning.

\section{Item Level Analysis}\label{app:Question}
Table~\ref{tab:score_quest} can be used to assess whether the AI struggles with the same items across different languages. To illustrate how to read Table~\ref{tab:score_quest}, let us consider the second most common language, Spanish. There were 416 items (with 1,248 responses) across all inventories in Spanish. For the corresponding items in English, there were 102 items with 0/3 correct, 42 items with 1/3 correct, 45 with 2/3 correct, and 227 with 3/3 correct in English. The percentages in the table show how the AI system performed on these items in Spanish. For the 102 items (306 responses) for which all three English responses were incorrect, 56 of the responses, corresponding to 18\%, were correct in Spanish. Hence, items that were extremely difficult in English (no correct answers at all) were also difficult in Spanish. This increases to 32\% (40/126) for the 42 items with 1/3 correct in English, 56\% (75/135) for the 45 items with 2/3 correct in English, and 88\% (601/681) for the 227 items where all three English responses were correct. The data show that as the items become easier in English, they also become easier in Spanish. Similar trends can be observed for other languages. The only notable exceptions results for Punjabi and Tamil, which have only 20\% and 26\% correctness for the items where the AI performed well in English (3/3).\\

\begin{table}
\caption{ Percentage of correct AI responses in various languages, grouped by the corresponding difficulty level of the same items in English. Items are categorized based on how many English responses (out of three) were correct. Each cell shows the percentage of correct responses in the given language, with the number of items for each category provided in parentheses.}
\begin{ruledtabular}
\begin{tabular}{lcccc}
&\multicolumn{4}{c}{Answers correct in English}\\
Language & 0/3 & 1/3 & 2/3 & 3/3 \\
\hline
Arabic & 17\% (16) & 8\% (4) & 50\% (6) & 85\% (50) \\
Bengali & 5\% (7) & 0\% (2) & 11\% (3) & 61\% (18) \\
Catalan & 14\% (7) & 0\% (2) & 78\% (3) & 93\% (18) \\
Chinese & 21\% (46) & 43\% (24) & 40\% (15) & 73\% (153) \\
Croatian & 3\% (11) & 33\% (6) & 67\% (4) & 92\% (42) \\
Czech & 17\% (8) & 8\% (4) & 76\% (7) & 88\% (33) \\
Dutch & 14\% (7) & 0\% (2) & 56\% (3) & 89\% (18) \\
Finnish & 24\% (33) & 48\% (22) & 48\% (11) & 90\% (85) \\
Filipino & 10\% (7) & 17\% (2) & 44\% (3) & 85\% (18) \\
French & 23\% (42) & 42\% (12) & 43\% (21) & 89\% (86) \\
German & 14\% (26) & 36\% (15) & 41\% (9) & 87\% (46) \\
Greek & 14\% (36) & 43\% (17) & 52\% (9) & 83\% (49) \\
Hebrew & 10\% (7) & 17\% (2) & 67\% (3) & 67\% (18) \\
Hindi & 14\% (7) & 0\% (2) & 11\% (3) & 74\% (18) \\
Hungarian & 19\% (7) & 17\% (2) & 56\% (3) & 93\% (18) \\
Indonesian & 25\% (29) & 48\% (16) & 44\% (15) & 84\% (107) \\
Icelandic & 14\% (7) & 0\% (2) & 67\% (3) & 93\% (18) \\
Italian & 17\% (18) & 22\% (6) & 56\% (6) & 92\% (26) \\
Japanese & 16\% (47) & 31\% (14) & 35\% (17) & 83\% (163) \\
Malay & 9\% (29) & 37\% (10) & 52\% (9) & 82\% (40) \\
Norwegian & 10\% (7) & 33\% (2) & 56\% (3) & 98\% (18) \\
Persian & 20\% (18) & 33\% (6) & 28\% (6) & 59\% (26) \\
Punjabi & 10\% (7) & 0\% (2) & 56\% (3) & 20\% (18) \\
Polish & 29\% (7) & 17\% (2) & 67\% (3) & 98\% (18) \\
Portuguese & 26\% (44) & 37\% (20) & 48\% (14) & 84\% (125) \\
Russian & 5\% (7) & 17\% (2) & 67\% (3) & 78\% (18) \\
Serbian & 67\% (1) & 33\% (3) & 0\% (1) & 70\% (19) \\
Slovak & 14\% (7) & 17\% (2) & 67\% (3) & 85\% (18) \\
Spanish & 18\% (102) & 32\% (42) & 56\% (45) & 88\% (227) \\
Swedish & 18\% (49) & 36\% (29) & 44\% (18) & 84\% (122) \\
Tamil & 5\% (7) & 50\% (2) & 22\% (3) & 26\% (18) \\
Thai & 17\% (21) & 25\% (8) & 24\% (7) & 59\% (76) \\
Turkish & 15\% (41) & 13\% (13) & 45\% (11) & 84\% (31) \\
Ukrainian & 17\% (10) & 0\% (2) & 33\% (1) & 74\% (13) \\
\hline
Overall &  &  &  & \\
non-English & 18\% (725) & 33\% (301) & 48\% (274) & 82\% (1771)\\
\hline
Overall &  &  &  & \\
non-English &  &  &  & \\
Text only & 16\% (51) & 52\% (18) & 65\% (51) & 84\% (759)\\
\end{tabular}
\end{ruledtabular}
\label{tab:score_quest}
\end{table}

Table~\ref{tab:incorrect_Q} shows the items that had two or three incorrect answers for the set of the three AI responses. It then shows if the selected multiple-choice items were the same or different. The percentages for the full data set as well as the data set split into English and non-English responses show that the AI frequently picked the same incorrect multiple-choice items.\\

\begin{table}
\caption{Incorrect answer analysis. Percentages with the corresponding number of items in parenthesis for which the incorrect answers were the same or different.}
\begin{ruledtabular}
\begin{tabular}{llccc}
Incorrect & different & &  &  \ \\
answers & or same & All &  English &  Non-English \\
\hline
3 & 3 different &  8\% (98) &   6\% (18) &  9\% (80)\\
3 & 3 the same  &   53\% (617) & 61\% (173) &  50\% (444)\\
3 & 2 the same &   38\% (448) & 33\% (93)&   40\% (355) \\
\hline
2 & 2 different &   34\% (197)   &28\% (37) &   36\% (160)\\
2& 2 the same &   66\% (376) &   72\% (94) &   64\% (282)\\
\end{tabular}
\end{ruledtabular}
\label{tab:incorrect_Q}
\end{table}

If picking a particular incorrect answer was random, the theoretical probabilities would be different from the percentages in Table~\ref{tab:incorrect_Q}. Hence, the AI system gravitates toward particular incorrect answers. Here are the calculations to get to those probabilities:
\begin{itemize}
\item When two of the three answers are incorrect, we can have two possible outcomes.

For the two incorrect answers to be the same, we need to choose which of the 4 incorrect options appears twice, giving us 4 possibilities. The total number of ways two incorrect answers can be the same is therefore 4, while the total possible combinations of 2 incorrect answers from 4 options is $4 \times 4 = 16$. This gives us a probability of $4/16 = 1/4 = 0.25$.
For the two incorrect answers to be different, the first incorrect answer can be any of the 4 options, and the second incorrect answer must be different from the first (3 possibilities). The total number of ways this can happen is $4 \times 3 = 12$, and the total possible combinations remains 16, giving us a probability of $12/16 = 3/4 = 0.75$.

\item When all three picks are incorrect, we have three possible scenarios.

For all three incorrect answers to be the same, we need to pick the same incorrect option three times. For any specific incorrect option, the probability is $(1/4) \times (1/4) \times (1/4) = 1/64$. Since we have 4 different incorrect options to choose from, the probability becomes $4 \times (1/64) = 4/64 = 1/16 = 0.0625$.

For all three incorrect answers to be different, the first incorrect answer can be any of the 4 options, the second must be different from the first (3 possibilities), and the third must be different from both previous picks (2 possibilities). The total number of ways this can happen is $4 \times 3 \times 2 = 24$. With total possible outcomes when picking from 4 options three times being 4³ = 64, the probability is $24/64 = 6/16 = 0.375$.

Finally, for exactly two incorrect answers to be the same, we can have three patterns: ($1=2\ne3$), ($1=3\ne2$), or ($1\ne2\ne3$). For each pattern, we choose which of 4 options is repeated (4 possibilities) and which option appears for the non-repeating position (3 possibilities), giving $4 \times 3 = 12$ ways for each pattern. The total ways across all three patterns is $12 + 12 + 12 = 36$, resulting in a probability of $36/64 = 9/16 = 0.5625$.\\
\end{itemize}

\end{document}